\documentclass[11pt,a4paper]{article}
\usepackage{jcappub}
\usepackage{lmodern}
\usepackage{graphicx,color}
\graphicspath{{figures/}}
\usepackage{comment}
\usepackage{cancel}

\usepackage[T1]{fontenc}
\usepackage[utf8]{inputenc}
\usepackage{color}
\usepackage{amsmath}
\usepackage{amssymb}
\usepackage{esint}
\makeatletter

\definecolor{somegreen}{cmyk}{0,0.49,0.98,0.09}
\definecolor{red}{rgb}{1,0,0}
\definecolor{magenta}{cmyk}{0,1,0,0}
\definecolor{lavender}{cmyk}{0.16,0.67,0,0.57}
\definecolor{darkgreen}{rgb}{0,0.65,0.05}
\definecolor{antiquefuchsia}{rgb}{0.33, 0.1, 0.89}

\newcommand{\dd}{\textrm{d}}

\let\jnl@style=\rm
\def\ref@jnl#1{{\jnl@style#1}}

\def\aj{\ref@jnl{AJ}}                   % Astronomical Journal
\def\actaa{\ref@jnl{Acta Astron.}}      % Acta Astronomica
\def\araa{\ref@jnl{ARA\&A}}             % Annual Review of Astron and Astrophys
\def\apj{\ref@jnl{ApJ}}                 % Astrophysical Journal
\def\apjl{\ref@jnl{ApJ}}                % Astrophysical Journal, Letters
\def\apjs{\ref@jnl{ApJS}}               % Astrophysical Journal, Supplement
\def\ao{\ref@jnl{Appl.~Opt.}}           % Applied Optics
\def\apss{\ref@jnl{Ap\&SS}}             % Astrophysics and Space Science
\def\aap{\ref@jnl{A\&A}}                % Astronomy and Astrophysics
\def\aapr{\ref@jnl{A\&A~Rev.}}          % Astronomy and Astrophysics Reviews
\def\aaps{\ref@jnl{A\&AS}}              % Astronomy and Astrophysics, Supplement
\def\azh{\ref@jnl{AZh}}                 % Astronomicheskii Zhurnal
\def\baas{\ref@jnl{BAAS}}               % Bulletin of the AAS
\def\bac{\ref@jnl{Bull. astr. Inst. Czechosl.}}
\def\caa{\ref@jnl{Chinese Astron. Astrophys.}}
\def\cjaa{\ref@jnl{Chinese J. Astron. Astrophys.}}
\def\icarus{\ref@jnl{Icarus}}           % Icarus
\def\jcap{\ref@jnl{J. Cosmology Astropart. Phys.}}
\def\jrasc{\ref@jnl{JRASC}}             % Journal of the RAS of Canada
\def\memras{\ref@jnl{MmRAS}}            % Memoirs of the RAS
\def\mnras{\ref@jnl{MNRAS}}             % Monthly Notices of the RAS
\def\na{\ref@jnl{New A}}                % New Astronomy
\def\nar{\ref@jnl{New A Rev.}}          % New Astronomy Review
\def\pra{\ref@jnl{Phys.~Rev.~A}}        % Physical Review A: General Physics
\def\prb{\ref@jnl{Phys.~Rev.~B}}        % Physical Review B: Solid State
\def\prc{\ref@jnl{Phys.~Rev.~C}}        % Physical Review C
\def\prd{\ref@jnl{Phys.~Rev.~D}}        % Physical Review D
\def\pre{\ref@jnl{Phys.~Rev.~E}}        % Physical Review E
\def\prl{\ref@jnl{Phys.~Rev.~Lett.}}    % Physical Review Letters
\def\pasa{\ref@jnl{PASA}}               % Publications of the Astron. Soc. of Australia
\def\pasp{\ref@jnl{PASP}}               % Publications of the ASP
\def\pasj{\ref@jnl{PASJ}}               % Publications of the ASJ
\def\rmxaa{\ref@jnl{Rev. Mexicana Astron. Astrofis.}}%
\def\qjras{\ref@jnl{QJRAS}}             % Quarterly Journal of the RAS
\def\skytel{\ref@jnl{S\&T}}             % Sky and Telescope
\def\solphys{\ref@jnl{Sol.~Phys.}}      % Solar Physics
\def\sovast{\ref@jnl{Soviet~Ast.}}      % Soviet Astronomy
\def\ssr{\ref@jnl{Space~Sci.~Rev.}}     % Space Science Reviews
\def\zap{\ref@jnl{ZAp}}                 % Zeitschrift fuer Astrophysik
\def\nat{\ref@jnl{Nature}}              % Nature
\def\iaucirc{\ref@jnl{IAU~Circ.}}       % IAU Cirulars
\def\aplett{\ref@jnl{Astrophys.~Lett.}} % Astrophysics Letters
\def\apspr{\ref@jnl{Astrophys.~Space~Phys.~Res.}}
\def\bain{\ref@jnl{Bull.~Astron.~Inst.~Netherlands}}
\def\fcp{\ref@jnl{Fund.~Cosmic~Phys.}}  % Fundamental Cosmic Physics
\def\gca{\ref@jnl{Geochim.~Cosmochim.~Acta}}   % Geochimica Cosmochimica Acta
\def\grl{\ref@jnl{Geophys.~Res.~Lett.}} % Geophysics Research Letters
\def\jcp{\ref@jnl{J.~Chem.~Phys.}}      % Journal of Chemical Physics
\def\jgr{\ref@jnl{J.~Geophys.~Res.}}    % Journal of Geophysics Research
\def\jqsrt{\ref@jnl{J.~Quant.~Spec.~Radiat.~Transf.}}
\def\memsai{\ref@jnl{Mem.~Soc.~Astron.~Italiana}}
\def\nphysa{\ref@jnl{Nucl.~Phys.~A}}   % Nuclear Physics A
\def\physrep{\ref@jnl{Phys.~Rep.}}   % Physics Reports
\def\physscr{\ref@jnl{Phys.~Scr}}   % Physica Scripta
\def\planss{\ref@jnl{Planet.~Space~Sci.}}   % Planetary Space Science
\def\procspie{\ref@jnl{Proc.~SPIE}}   % Proceedings of the SPIE

\makeatother

\title{Fisher matrix for the one-loop galaxy power spectrum:
measuring expansion  and growth rates without assuming a cosmological model
}
\author[a]{Luca Amendola,}
\author[b]{Massimo Pietroni,}
\author[a,c,d]{and Miguel Quartin}
\affiliation[a]{Institute of Theoretical Physics, Philosophenweg 16, Heidelberg University, 69120, Heidelberg, Germany}
\affiliation[b]{Department of Mathematical, Physical and Computer Sciences, University of Parma, and INFN, Gruppo Collegato di Parma, Parco Area delle Scienze 7/A, 43124, Parma, Italy}
\affiliation[c]{Instituto de Física, Universidade Federal do Rio de Janeiro, 21941-972, Rio de Janeiro, RJ, Brazil}
\affiliation[d]{Observatório do Valongo, Universidade Federal do Rio de Janeiro, 20080-090, Rio de Janeiro, RJ, Brazil}
\date{\today }

\abstract{We introduce a methodology to extend the Fisher matrix forecasts to mildly non-linear scales without the need of selecting a cosmological model. We make use of standard non-linear perturbation theory for biased tracers complemented by  counterterms, and assume that the cosmological distances can be measured accurately with standard candles. Instead of choosing a specific model, we parametrize the linear power spectrum and the growth rate in several $k$ and $z$ bins.  We show that one can then obtain model-independent constraints of the expansion rate $E(z)=H(z)/H_0$ and the growth rate $f(k,z)$, besides the  bias functions. We apply the technique to both Euclid and DESI public specifications in the range $0.6\le z \le 1.8$ and show that the gain in precision when going from $k_{\rm max} = 0.1$ to $0.2\,h$/Mpc is around two- to threefold, while it reaches four- to ninefold when extending to $k_{\rm max} = 0.3\,h$/Mpc. In absolute terms, with $k_{\rm max}=0.2\,h/$Mpc, one can reach high precision on $E(z)$ at each $z$-shell: 8--10\% for DESI with $\Delta z=0.1$, 5--6\% for Euclid with $\Delta z=0.2-0.3$. This improves to 1--2\% if the growth rate $f$ is taken to be $k$-independent. The  growth rate itself has in general much weaker constraints, unless assumed to be $k$-independent, in which case the gain is similar to the one for $E(z)$ and uncertainties around 5--15\% can be reached at each $z$-bin. We also discuss how  neglecting the non-linear corrections can have a large effect on the constraints even for $k_{\rm max}=0.1\,h/$Mpc, unless one has independent strong prior information on the non-linear parameters. }

\begin{document}
\maketitle

\section{Introduction}

Large-scale galaxy surveys are providing a better
and better understanding of cosmology. In the near future, both ground surveys like   DESI~\citep{DESI:2016fyo}, 4MOST~\citep{4MOST2019}, J-PAS~\citep{Bonoli:2020ciz} and the Rubin Observatory Legacy Survey of Space and Time (LSST)~\citep{LSSTScience:2009jmu} and space surveys such as Euclid~\citep{laureijs2011euclid,Amendola:2016saw} and the Nancy Roman Space Telescope survey (Roman)~\citep{Eifler:2020hoy,Rose:2021nzt} will expand our spatial and temporal knowledge by orders of magnitude, promising sub-percent estimates of the main cosmological parameters. Together with other sources of information, from Hubble diagrams to CMB, the possibility of finding  deviations from the standard cosmological model, or its definitive confirmation, seems within reach.

One of the main difficulties in the data analysis of galaxy clustering is the optimal exploiting of the non-linear information for cosmological purposes. Scales below some tens of Megaparsecs are  subject to non-linear evolution, but an accurate modelling of this evolution is hard to produce. In the last decades valiant efforts were put forward in developing both semi-analytical methods, improving standard perturbation theory~\cite{Bernardeau_2002} via resummation techniques~\cite{Crocce:2005xy,Matarrese:2007wc,Taruya2007,Pietroni08,Bernardeau:2008fa,Senatore:2014via,Baldauf:2015xfa} and effective field theory treatment of the effects of the short scales on the intermediate ones \cite{Baumann:2010tm,Pietroni:2011iz,Carrasco:2012cv,Manzotti:2014loa},  and cosmological simulations, which often rely only on gravitational effects (for a recent review see \cite{Angulo:2021kes}) but more recently also on more sophisticated hydrodynamical implementations (see e.g.~\citep{Khandai:2014gta,Crain:2015poa,McCarthy:2016mry,Dave:2019yyq,Castro:2020yes,Barreira:2021ukk}). Nevertheless, these methods have been in fact developed and tested only for standard cosmologies and a few other selected cases. A restriction to a limited set of cosmologies carries the risk of missing new physics and of biasing the parameter estimation when analysing real data.

In this paper we perform forecasts for future surveys following  a complementary route. Instead of focusing on specific cosmological models, we develop a methodology to extract from the data information that is as much model-independent as possible, following the approach already presented at the linear level in \cite{2020JCAP...11..054B,Amendola:2019lvy,Quartin:2021dmr} (see also \cite{Samushia:2010ki} for an earlier work). More specifically, we do not assume a cosmological model neither at the background nor at the perturbed level, and parametrize galaxy/matter biasing on general grounds, via the perturbative bias expansion, which assumes only the equivalence principle (for a review, see \cite{Desjacques:2016bnm}). As we discuss in detail in Appendix E, in our approach the signal comes essentially from the Alcock-Paczyński effect and is based, ultimately, on the assumption of statistical isotropy. We still need to assume, however, also a homogeneous and isotropic background with relatively small perturbations that can be modelled up to second order.

We use a Fisher matrix analysis extended, for the first time, to  one-loop level for biased tracers, in which the parameters are not the usual cosmological ones, but rather  the linear power spectrum and the linear growth rate in wavenumber and redshift bins, plus a combination of $H(z)$ and of distance, plus four free functions of bias up to second order, and one or more ``counterterm'' parameters. These elements combine into the non-linear one-loop power spectrum and constitute our theory-informed parametrization for the mildly non-linear scales.

The investigation of this paper is still preliminary because the theoretical one-loop power spectrum that we employ has been so far only tested in a limited number of cases beyond $\Lambda$CDM. It might well be therefore that our parametrization, even if vastly more general than those based on specific cosmological models, turns out to be still insufficient to reproduce with high fidelity the non-linear behaviour. The method we describe here, however, can be directly improved to more general forms that might be developed in the future, for instance using the general forms for the perturbative kernels derived in~\cite{DAmico:2021rdb}. Therefore, we believe that, notwithstanding its current limitations, our method is a useful step forward.

We show that the non-linear corrections allow to reconstruct two fundamental cosmological functions: the product $H(z)D(z)$, where $D(z)$ is the cosmological (either luminosity or angular diameter) distance and the linear perturbation growth rate $f(k,z)$ (including its possible $k$-dependence). Provided that $D(z)$ can be accurately measured by e.g. the Roman Space Telescope (see \cite{Rose:2021nzt}) with Type Ia supernovae, or other standardized source, we can therefore extract $H(z)$.
More exactly, while standard sirens are capable of measuring directly $D$~\cite{Schutz:1986gp}, supernovae instead measure only $H_0 D$. So with supernovae we can derive only $E(z)=H(z)/H_0$: in this case,  $H(z)$  can only be measured up to the uncertainty in $H_0$ at the time.

Other approaches have been considered in the literature to measure $H(z)$ with galaxy clustering with different degrees of model-independence. Since radial BAO measures $H(z) r_d$, where $r_d$ is the primordial sound horizon, $H(z)$ could be measured with BAO from the so-called inverse distance ladder if a prior on $r_d$ could be justified~\cite{Cuesta:2014asa}. This however relies on models for the early universe physics.  To circumvent this one could use a low redshift anchor, to wit local measurements of $H_0$ to recover $H(z)$ as performed by~\cite{Bernal:2016gxb}. In either case, one has to assume that the tracers' spectral shape  does not bias the location of the baryonic wiggles. More recently, the ${\alpha_\parallel,\alpha_\perp,f\sigma_8}$ parametrization became an established procedure to improve model-independence (see e.g. \cite{Zhao:2015gua, BOSS:2016wmc, BOSS:2016hvq, Foroozan:2021zzu}), but it still relies on assumptions regarding the shape of $P(k)$ and of  $\beta(k)$ (typically taken to be $k$-independent). The method we employ here is an extension of this approach to allow a completely free shape for $P(k),\beta(k)$.

It is important to remark that $f(k,z)$ cannot be measured in a model-independent way within linear scales because it is fully degenerate with the linear bias parameter $b_1(z)$: only their combination $\beta(k,z) = f/b_1$, the  redshift distortion parameter, enters the power spectrum equations.

Beside $E(z)$ and $f(k,z)$, we can obtain constraints also on the  bias and counterterm parameters, as well as the power spectrum itself. We apply the method to surveys that approximate the expected specifications of the Euclid and DESI galaxy surveys,\footnote{Although for brevity we refer often to   Euclid and DESI surveys, in both cases it is understood that we are not providing official specifications but just use publicly available information.} covering the redshift range from 0.6 to 1.8. We also compare our method to the constraints provided by~\cite{Ivanov:2019pdj} for the BOSS data (similar analyses have been performed in~\cite{DAmico:2019fhj,Colas:2019ret,Philcox:2020vvt,Chen:2021wdi}): although several differences in the approaches prevent a close match, we find overall a reasonable agreement.

The bottom line of this paper is in line with several previous works: the advantage of going to even mildly non-linear scales is large. We find that if it is possible to accurately extend the data analysis from $k_{\rm max}=0.1\,h/$Mpc to $k_{\rm max}=0.2\,h/$Mpc, one stands to gain a factor of roughly three  in every redshift bin in the constraints for $E(z)$ and $f(k,z)$ for both Euclid and DESI. Much better constraints, down to 1--2\% are obtained  if one assumes $f$ to be $k$-independent. An extension to $k_{\rm max}=0.3\,h/$Mpc, although probably still unwarranted by current modelling, would increase the gain by a factor up to nine.
For the growth rate $f$, however, we find that, notwithstanding the gain, the uncertainties remain large, typically above 20\%. When $f$ is taken to be $k$-independent, however, we find that  an uncertainty smaller than 10\% can be achieved by adopting an upper cut-off $k_{\rm max}=0.3\,h/$Mpc.

As another interesting result of this work,  we  find that neglecting the non-linear corrections can have a large effect on the constraints even for a cut-off at $k=0.1\,h/$Mpc, as often employed in literature, unless one has independent strong prior information on the non-linear parameters.

We remark that in practice the determination of $k_{\rm max}$ must be made with the use of simulations in order to see at which scales parameter reconstruction start to become biased. One possibility is through blind challenges~\cite{Nishimichi:2020tvu,Brieden:2022ieb}. In any case, such a procedure is beyond the scope of this work, but it is a necessary follow-up before real data can be analysed robustly.

\section{Power spectrum at one loop}

We will adopt a model for the galaxy power spectrum based on one-loop perturbation theory augmented by UV counterterms, shot noise terms, and a smoothing factor that models spectroscopic errors.  It can be written as
\begin{align}
    P_{gg}(k,\mu,z) & =S_{\rm g}(k,\mu, z )^2 \left[ P^{\rm lin}(k,\mu, z)+P^{\rm 1loop}(k,\mu, z)+P^{\rm UV}(k,\mu,z)\right] + P^{\rm sn}(z)\,,
    \label{Pgg}
\end{align}
where $k=(k_\parallel^2+k_\perp^2)^{1/2}$,   $\mu \equiv k_\parallel/k$, and $k_\parallel$ ($k_\perp$) is the component of the wavevector parallel (perpendicular) to the line of sight.
The linear contribution is given by
\begin{equation}
    P^{\rm lin}(k,\mu,z) = \left(1+ \mu^2 \beta(k,z)\right)^2 b_1(z)^2 P(k,z)\,,
\label{plin}
\end{equation}
where $P(k,z)$ is the linear matter power spectrum in real space, $b_1(z)$ is the linear bias parameter and $\beta(k,z)=f(k,z)/b_1(z)$, and $f(k,z) \equiv \dd \log \delta / \dd \log a$ is the linear growth rate. In the following, we will consider both the cases of a scale independent $\beta(z)$  and that of a fully scale-dependent one, free in each $k$ bin. Actually, our model Eq. (\ref{Pgg}) is derived under the assumption of a scale-independent growth. However, the present approach in which the linear power spectrum is a free parameter should be largely insensitive to the actual forms of the perturbation theory kernels and to a possible scale dependence of the bias parameters. Therefore, we show results also in the case of a scale-dependent linear growth, assuming that the leading scale-dependent effect is encoded in the growth function $f(k,z)$. For the same reason, we will use the Einstein-deSitter form for the perturbation theory kernels, assuming that the effect of their cosmology-dependence is largely reabsorbed by the bias coefficients. Indeed, as shown in \cite{DAmico:2021rdb}, typical variations of these kernels in different cosmologies are in the few percent range and will be probably largely degenerate with some of the parameters that we vary in our analysis. We plan, however, to come back to this issue in a future work.

The complete expression for $P^{\rm 1loop}(k,\mu, z)$, by now standard (see e.g. \cite{Ivanov:2019pdj, DAmico:2019fhj}), is given in Appendix~\ref{app:nonlin-P}. It contains, besides $b_1(z)$, three more tracer-dependent bias parameters, that we indicate with $b_2(z)$, $b_{{\cal{G}}_2}(z)$ and $b_{\Gamma_3}(z)$. Although in \cite{Ivanov:2019pdj} the last term was set to zero because almost degenerate with other terms, we decided to keep it because we find it not significantly more degenerate than other parameters. In Appendix~\ref{app:nonlin-P} we also discuss  the UV counterterms $P^{\rm UV}(k,\mu,z)$. We anticipate that our analysis, being  independent of the detailed power spectrum shape, turns out to be largely insensitive to the UV corrections compared to those in refs.~\cite{Ivanov:2019pdj, DAmico:2019fhj}.
For the same reason, we are largely insensitive to the exact location of the BAO wiggles, and therefore of the effect of bulk flows on the BAO wiggles, which translates in a damping factor on the oscillating part of the power spectrum, which we do not include in Eq.~(\ref{Pgg}). In any case, we test our method on a de-wiggled spectrum and show that the resulting constraints change by no more than 10\%.

Finally, we included the shot-noise spectrum
\begin{equation}
    P^{\rm sn}(z)=\frac{1}{ n(z)}(1+P_{\rm shot})\,,
\end{equation} where $n(z)$ is the density of the considered galaxies  at redshift $z$, and $P_{\rm shot}$ an additional parameter to be varied in the Fisher matrix.

The overall factor $S_{\rm g}(k,\mu, z )^2$ is a smoothing term that takes into account both the spectroscopic redshift errors and the Finger-of-God (FoG) effect. Following \cite{2020A&A...642A.191E,BOSS:2016psr}, we write
\begin{equation}
   S_{\rm g}(k,\mu, z )=\exp\left[-\frac{1}{2}(k\mu\sigma_{\rm z})^{2}\right]\exp\left[-\frac{1}{2}(k\mu\sigma_{f})^{2}\right],\label{eq:FoG}
\end{equation}
where,
\begin{equation}
    \sigma_{\rm z}=\sigma_0 (1+z)H(z)^{-1}\,.
\end{equation}
For the Euclid-like survey $\sigma_0=0.001$ (which, at $z\approx 1$,  corresponds to an effective smoothing over scales of $\approx 10$ Mpc$/h$), while for DESI the requirement is to have $\sigma_0 \simeq 0.0005$~\cite{DESI:2016fyo}. This value is small enough that for DESI we can assume $\sigma_0=0$ for simplicity. For the FoG smoothing factor
we assume
a fiducial $\sigma_{f}=5$Mpc/$h$.
Besides the exponential damping model, we investigate the EFTofLSS model for the FoG effect, via the second and last terms in eq.~\eqref{eq:UV-2}, which encode  the  effect of the short scale velocity modes on the large scale modes, at leading and at next-to-leading order, respectively (see for instance the discussion in \cite{Ivanov:2019pdj}). For both models, we will marginalize on the relative parameters, finding compatible results.

In all the momentum integrals, both for the 1-loop power spectrum and the Fisher matrix evaluation, we integrate over a smooth linear power spectrum generated with the CLASS Boltzmann code~\cite{Blas:2011rf}.

In the linear regime, only the combinations $\beta=f/b_1$ and $b_1^2 P$ entering Eq.~(\ref{plin}) can be measured, and not $f$, $b_1$ and $P$ separately.  So in the strictly linear regime using single tracers, $b_1$ cannot be used as a free parameter.
In contrast, it is important to remark that thanks to the nonlinear terms, we can measure $f$, $b_1$, and $P$ separately, regardless of cosmological parametrizations.

In summary,  the galaxy power spectrum $P_{gg}(k,\mu,z)$ would depend in general on four bias parameters ($b_1(z)$, $b_2(z)$, $b_{{\cal{G}}_2}(z)$, and $b_{\Gamma_3}(z)$) and three UV counterterms ($c_0(z)$, $c_2(z)$ and $\tilde c (z)$, see Appendix~\ref{app:nonlin-P}). All these parameters are coefficients of a gradient expansion (see \cite{Desjacques:2016bnm})  and therefore are function of $z$ alone and not of $k$. As already mentioned, instead of the counterterms $c_2,\tilde c$, we adopt the FoG parametrization of Eq.~\ref{eq:FoG}, but we also explore for comparison some cases with non-zero $c_2,\tilde c$. Our reference model contains then five new parameters (per $z$ bin) with respect to the linear regime, namely, $b_1(z)$, $b_2(z)$, $b_{{\cal{G}}_2}(z)$, $b_{\Gamma_3}(z)$ and $c_0(z)$.  We refer collectively to them  as NL parameters.

\section{Fisher matrix}

We see in Appendix~\ref{app:nonlin-P} that $P_{gg}$ depends on $P^{\rm lin}$ and $f$, which in turn depend on $k$ and $z$,  and on various other functions of $z$ alone that enter the correction terms. Since we want to remain as model-independent as possible, we do not parametrize the spectrum in terms of the usual cosmological parameters, which necessarily require a choice of model, but rather employ directly the data  in $z,k$ bins. We split the surveys in several $z$ bins that are assumed to be independent; therefore, from now on, we focus on the $k$ binning,  while the $z$-dependence of the various functions is always understood.

Quantities such as $k$, $\mu$, and volumes  must be computed assuming a reference model (subscript $r$). For any other cosmology, the correction, known the Alcock-Paczyński (AP) effect, is given by  $\mu=\mu_{r}\eta D_r/(D\alpha)$ and $k$ as $k=\alpha k_{r}$, where~\citep{2000ApJ...528...30M,Amendola:2004be,Samushia:2010ki}
\begin{equation}
    \alpha \,=\, \frac{D_r}{D} \sqrt{\mu_{r}^{2}(\eta^{2}-1)+1} \,,
    \label{eq:alpha}
\end{equation}
and
\begin{equation}
    \eta \,\equiv\, \frac{HD}{H_{r}D_{r}}  \,. \label{eq:eta}
\end{equation}
Moreover, all observed spectra  get multiplied by a volume-correcting factor $\Upsilon$~\citep{1996MNRAS.282..877B,Seo:2003pu}, so that $P_{\rm gg,\,obs} \rightarrow{\Upsilon P_{\rm gg}}$, where
\begin{equation}
     \Upsilon = \frac{H D_{r}^2}{H_{r}D^2} \,.
\end{equation}
This factor, however, is degenerate at the linear level with  the power spectrum amplitude itself in our model-independent method. The non-linear terms introduce some dependence on $\Upsilon$ but we estimate that it can be neglected so we do not consider it any further. Provided we measure $D$ precisely enough with usual standard candles methods (see below), the AP effect depends only on $\eta$, or equivalently on $H$. It is exactly because of this AP dependence that we can estimate $H$ independently of the cosmological model. Notice that $H(z)$ can be obtained by differentiating $D(z)$ only when assuming a flat space, which would then spoil our model-independent approach.

This procedure extends the well known ${\alpha_\parallel,\alpha_\perp,f\sigma_8}$ parametrization by first modelling nonlinearities with state of the art perturbation theory approaches (EFTofLSS and bias expansion), and secondly, by an enlarged Fisher matrix which includes at every redshift, besides $\eta$ and the bias and UV parameters, the linear $P(k)$ values and the redshift distortion function $\beta(k)$ at several different $k$-bins.

We can write now a FM for $P_{gg}$ that  depends for each $z$ bin on the set of parameters (collectively denoted as $\theta_\alpha$)
\begin{equation}
    \{\log P(k),\log \beta(k),\log \eta,\log b_{1}, b_{2}, b_{{\cal{G}}_2}, b_{\Gamma_3}, c_{0}, \log\sigma_f,P_{\rm shot} \} \,.
    \label{eq:parfull}
\end{equation}
Notice that we cannot take the $\log$ of some parameters because  they cannot be assumed to be positive. We choose to operate with the log of the positive-definite parameters so that the results are relative, rather than absolute, errors; this makes easier to compare results for the different cases. Since we take uninformative priors, the effect of taking the logarithm is very small. The same set of parameters is  varied independently at every  redshift.

The fiducial values for $P$ and $\beta$ are the standard $\Lambda$CDM ones. In particular, we adopt for $P(k)$ the following choice of cosmological parameters: $\Omega_c=0.270$, $\Omega_b=0.049$,  $\Omega_k=0$, $h=0.67$, $n_s=0.96$, and $\sigma_8=0.83$. Moreover, for $\beta(z)=f(z)/b$, we use the approximation $f(z)=\Omega_m(z)^\gamma$ with $\gamma=0.545$. For the shot-noise parameter $P_{\rm shot}$ the fiducial is 0. Finally, as already mentioned, the fiducial for $\sigma_f$ is 5 Mpc/$h$ as in~\cite{Ivanov:2019pdj}.

The choice of fiducial for the NL parameters is at this stage quite arbitrary since there are few measurements and no clear theoretical expectations. In this paper therefore we adopt tentatively the values obtained in \cite{Ivanov:2019pdj} for the BOSS survey (NGC, high-$z$, see their Table~10), except for $b_1$ for which we take an average of either the fiducial Euclid values  (from \cite{2020A&A...642A.191E}) or the DESI LRG+ELG average values in all redshifts, weighted by the expected number of objects and for $b_{\Gamma_3}$ which, as already mentioned, was not included in the analysis of \cite{Ivanov:2019pdj}  and for which we take a zero fiducial value:
\begin{equation}
\begin{aligned}
    &b_1=1.99 \;({\rm DESI}), \quad b_1=1.65 \;({\rm Euclid}), \\
    &b_2=-3.21,\quad b_{{\cal{G}}_2}=0.545,\quad b_{\Gamma_3}= 0,\\ &c_0=-53 \,({\rm Mpc}/h)^2\quad c_2=-21 \,({\rm Mpc}/h)^2\quad \tilde{c}=187 \,({\rm Mpc}/h)^4\,.
\end{aligned}
\end{equation}
These choices are of course not fully justified, since the BOSS survey samples different redshifts and  sources from both Euclid and DESI.  However, at this stage, any other choice of NL fiducials would be equally uncertain. To explore the dependence on fiducials, we also  experiment with different sets.

We now assume that the $N$ Fourier coefficients for the galaxy distribution $\delta_{\mathbf k}$ are Gaussian distributed with variance $P_{gg}({\mathbf k})$ (even if we know they are non-Gaussian at NL scales).
The FM for a set of parameters $\theta_{\alpha}$, in a survey of volume $V$, is then~\citep{Tegmark:1997rp,Abramo:2019ejj}
\begin{equation}
    F_{\alpha\beta} \,=\, \frac{1}{(2 \pi)^3} 2\pi k^{2}\Delta_{k}V\bar{F}_{\alpha\beta} \,=\, VV_{k}\bar{F}_{\alpha\beta}\,,
\end{equation}
where $V_{k}= (2\pi)^{-3}2\pi k^{2}\Delta_{k}$ is the volume of the Fourier space after integrating over the azimuthal angle but not over the polar angle (i.e., the volume of a spherical Fourier space shell of width $\Delta_{k}$ would be given by $\int d \mu \, V_k$). We defined as $\bar{F}$  the FM per unit phase-space volume $V V_k$ integrated over~$\mu$, i.e.
\begin{equation}
    \bar{F}_{\alpha\beta} = \frac{1}{2}\int_{-1}^{+1} d\mu \, \frac{\partial \ln P_{gg}}{\partial\theta_{\alpha}}\frac{\partial \ln P_{gg}}{\partial\theta_{\beta}} \,,
    \label{FMi}
\end{equation}
where the integrand is evaluated at the fiducial value.  We choose for simplicity a fixed number of $k$ intervals  between $k_{\rm min}$ and $k_{\rm max}$ so the $k$-bin sizes change with $k_{\rm max}$ ($k_{\rm min}$, which has very little impact on the results, is fixed to $0.001\,h/$Mpc).  We set the number of $k$-bins to $N=20$, equally spaced in $k^2$ space (in order to better sample  the high-$k$ end) and checked that the results do not change sensibly for higher $N$'s. In our reference case, therefore, for each $z$-shell, the parameter set $\theta_\alpha$ comprises  20 variables $\log P$, 20 variables $\log \beta$, and eight $k$-independent parameters.

We need then to take derivatives of $P_{gg}$ with respect to each of the parameters. These are discussed  in detail in  Appendix~\ref{app:der}.

The power spectrum covariance implied by our Fisher Matrix analysis is diagonal but non gaussian, as its diagonal components are proportional to the square of the nonlinear PS, not the linear one. However, at the perturbative order we are working, other non gaussian contributions both in the diagonal and the  non diagonal terms, are induced by the trispectrum, the survey geometry, and super sample variance. The use of analytical covariances was recently studied in great detail in \cite{Wadekar:2019rdu} and \cite{Wadekar:2020hax}, where it was found that the effect of non gaussian terms on parameter constraints, after marginalization over cosmological and bias parameters, is typically less than 10 \% (see for instance Fig. 1 in \cite{Wadekar:2020hax}).

\subsection{Estimate of the error on the distance}

In our method we can also estimate directly $D$,  since the AP effect on $k$ depends on $D$ as well (see Eq. \ref{eq:alpha}). However,  we find that the error is quite large, typically substantially more than 10\%, and therefore we need an external estimate of $D$ to convert the uncertainty on $\eta$ into an uncertainty on $H$.

We can estimate the relative error on  $D$ from supernovae Hubble diagrams in a given redshift bin as follows. For a survey with $N$ Type Ia supernovae in a $z$ bin, each with total statistical magnitude uncertainty $\sigma_{\rm int}$, the relative error in $\hat D=H_0 D$ is
\begin{equation}
    \frac{\Delta \hat D}{\hat D} = \frac{\log10}{5}\frac{\sigma_{\rm int}}{\sqrt{N}} \,.
\end{equation}
Assuming $\sigma_{\rm int}=0.13$,  one reaches, say, 0.1\% with  3500 supernovae.\footnote{Systematical errors on the other hand cannot be so easily dealt with. Here we assume they are negligible for simplicity, but they may well be the dominant source of uncertainties for large $N$.}
According to~\cite{Quartin:2021dmr} for $z \lesssim 0.5$  the performance of LSST would produce $N$ large enough to be more than sufficient.
 Both DESI and Euclid, however, will observe most of their galaxies at higher redshifts, where LSST detections are more limited. So beyond this redshift, we can adopt the estimates provided in \cite{Rose:2021nzt} for the Roman Space Telescope (see Table~\ref{tab:survey-specs}), which includes various sources of error beside $\sigma_{\rm int}$ (but again not systematic ones), and which are always smaller than 0.4\%.
Notice that these estimates are model-independent in the same sense as we employ here, namely they do not depend on a cosmological model.
 Adding this uncertainty in quadrature to $\eta$ we can obtain the final relative  error  on $E$:
\begin{equation}
    \left(\frac{\Delta E}{E}\right)^2=\left(\frac{\Delta \hat D}{\hat D}\right)^2+\left(\frac{\Delta\eta}{\eta}\right)^2. \label{eq:quad}
\end{equation}
In practice, however, the uncertainty induced by $\hat D$ is always much smaller than the one we obtain here for $\eta$ as long as $k_{\rm max}\le 0.3 \,h/$Mpc, so we can safely neglect this correction at this stage.

\begin{table}
    \centering
    %\footnotesize
    \setlength{\tabcolsep}{6.7pt}
    \begin{tabular}{ccccc}
    \hline
    $z$ & $V ({\rm Gpc}/h)^{3}$ & $10^{3}\cdot n  (h/{\rm Mpc})^3$ & $N_{\rm SN}$  & $\Delta \hat D/\hat D$ \\
    \hline
    \multicolumn{5}{c}{DESI (LRG+ELG)}\\
    \hline
    0.6--0.7 & 2.43 & 0.657 & 759  & 0.00384 \\
    0.7--0.8 & 2.89 & 1.58  & 871  & 0.00359 \\
    0.8--0.9 & 3.31 & 1.09  & 1010 & 0.00333 \\
    0.9--1.0 & 3.69 & 0.897 & 1119 & 0.00316 \\
    1.0--1.1 & 4.03 & 0.518 & 1167 & 0.00310 \\
    1.1--1.2 & 4.32 & 0.444 & 1100 & 0.00319 \\
    1.2--1.3 & 4.57 & 0.410 & 1021 & 0.00331 \\
    \hline
    \multicolumn{5}{c}{Euclid}\\
    \hline
    0.9--1.1 & 7.94 & 0.686  & 2285 & 0.00221  \\
    1.1--1.3 & 9.15 & 0.558  & 2121 & 0.00230  \\
    1.3--1.5 & 10.1 & 0.421  & 1687 & 0.00258  \\
    1.5--1.8 & 16.2 & 0.261  & 1130 & 0.00315  \\
    \hline
    \end{tabular}
    \caption{\label{tab:survey-specs} Survey specifications assumed in this work (see \cite{2020A&A...642A.191E} and \cite{Vargas-Magana:2018rbb}). We also list the expected number $N_{\rm SN}$ of supernovae Ia using the Roman Space Telescope and the relative distance errors $\Delta \hat D/\hat D$ in each bin  (see \cite{Rose:2021nzt}).
    }
\end{table}

\section{Comparison with BOSS and Forecasts for Euclid and DESI}

In order to proceed, we must determine the $k_{\rm max}$ at which we cut the power spectrum. In an analysis on data from  simulations, this can be done by increasing $k_{\rm max}$ up to the point at which the shifts in the central values of the posterior distributions of the cosmological parameters become significant (see for instance \cite{Nishimichi:2020tvu,Brieden:2022ieb}). Besides the redshift and volume of the considered survey, the result depends on the specific theoretical model. For instance for the analysis of BOSS data  \cite{Chudaykin:2020ghx} and \cite{Ivanov:2019pdj}  employed $k_{\rm max}=0.20 \,h$/Mpc and  $k_{\rm max}= 0.25
\,h$/Mpc, respectively. Other authors claim that even $k_{\rm max}=0.35 \,h$/Mpc may result in unbiased constrains~\cite{Eggemeier:2020umu}.
Moreover, since non-linearities are less important at higher redshifts, it is expected that $k_{\rm max}$ should naturally increase as a function of redshift~\cite{Nishimichi:2008ry,Tomlinson:2022xud}.

Therefore, regarding the present forecast, we consider as our baseline three maximal values for $k$: 0.1, 0.2 and 0.3$\,h/$Mpc which we apply independently for each redshift bin. The first one is the value often chosen to cut out the NL effects, which however are not really negligible even in $\Lambda$CDM  and might be more important in other cosmological models. The higher cut, $k_{\rm max}=0.3\,h/$Mpc is perhaps too optimistic, although  at the high redshifts of Euclid and in our shape-independent method even such high $k_{\rm max}$ might be acceptable. The central value $k_{\rm max}=0.2\,h/$Mpc represents probably the best compromise between  accuracy and precision, and constitute our ``reference'' case in the following.

\begin{figure}
    \centering
    \includegraphics[width=1.0\columnwidth]{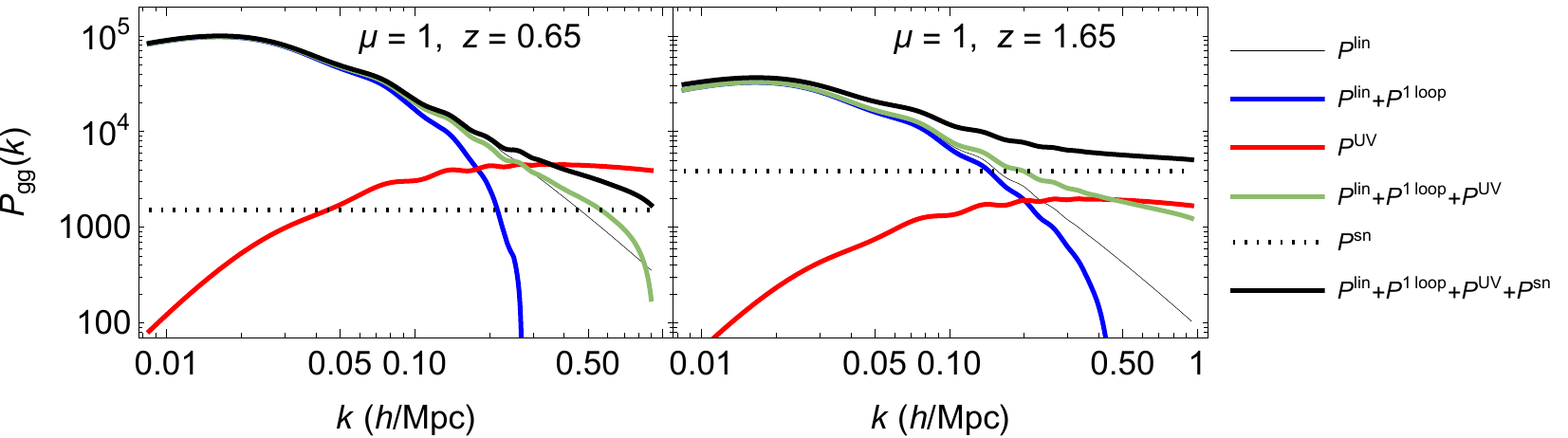}
    \caption{Galaxy power spectra for $\mu=1$ at $z=0.65$ (DESI) and $z=1.65$ (Euclid), without FoG corrections.
    \label{fig:pkDandE}}
\end{figure}

 We display some representative power spectra for Euclid and DESI in Figure~\ref{fig:pkDandE}. At high $z$ and high $k$'s the shot noise dominates the signal and we expect therefore, in these regimes, a weakening of the constraints.

\subsection{Comparison with BOSS}
Before producing forecasts, we compare our FM results with the real-data analysis of BOSS data \cite{Ivanov:2019pdj}. The two approaches are actually quite different, since we do not parametrize the power spectrum in terms of a cosmological model.  Moreover, our Fisher method forces us to use Gaussian priors, while in Ref.~\cite{Ivanov:2019pdj} flat priors are employed. Therefore, we cannot expect a close agreement. Still, we think it is interesting to perform this comparison to test how much the constraints on the NL parameters   depend  on the power spectrum parametrization. Moreover, we would like to quantify the constraints one could get from current BOSS data on $E$ and $f$ by using our method.

For this comparison, we  adopt the same Ref.~\cite{Ivanov:2019pdj} counterterms (see Eq.~\ref{eq:UV-2}), fiducials, and priors (equalling our Gaussian variance with the variance of the flat priors assumed in that paper), their cut at $k_{\rm max}=0.25 h/$Mpc, and the assumption that $\beta$ does not depend on $k$. Doing so we find that the uncertainties on the NL parameters are comparable, see  Table~\ref{tab:BOSS}. Notice that in \cite{Ivanov:2019pdj} the constraints on the spectral amplitude $A$ and on ${\tilde b}_i\equiv b_i\sqrt{A}$  (with $i=1,\,2$ or ${\cal{G}}_2$) are given. We derive the relative constraints on $b_i$ by simple quadrature, $\sigma_i^2=\sigma^2_{\tilde b_i}+\sigma_A^2/4$, where the errors are the relative ones (converted to absolute ones for $b_2,b_{{\cal{G}}_2} $). This exercise shows that from BOSS data, and adopting the priors in \cite{Ivanov:2019pdj}, we could get an uncertainty of approximately 2\% on $E(z)$ and of 12$\div$13\% on $f$ on both surveys.

\begin{table}
    \centering
    \begin{tabular}{cccccccccc}
    \hline
    \multicolumn{10}{|c|}{$k_{\mathrm{max}}=0.25\,h/$Mpc}\tabularnewline
    \hline &
    $z$ & $\Delta c_{0}$ &  $\Delta c_{2}$ &  $\Delta \tilde{c}$ & ${\Delta b_{1} / b_{1}}$
    & $\Delta b_{2}$
    & $\Delta b_{{\cal{G}}_2}$ & $\frac{\Delta E}{E}=\frac{\Delta\eta}{\eta}$ & ${\Delta f / f}$
    \tabularnewline
    \hline
    \hline
    this paper & 0.38  & 12 & 18 & 389 & 0.16 & 0.60 &  0.31 & 0.028 & 0.12  \\
    Ivanov et al.       & 0.38  & 31 & 35 & 240 & 0.11 &  0.81 & 0.38 & -& - \\
    \hline
    this paper & 0.61 & 23 & 13 & 223 & 0.12 & 0.72 & 0.46 & 0.017 & 0.09  \\
    Ivanov et al.  & 0.61  &  52 & 51 & 123 & 0.11 &  1.1 &  0.57 & - & -\\
    \hline
    \end{tabular}
    \caption{\label{tab:BOSS}   Comparison with Ref. \cite{Ivanov:2019pdj} for the  NGC assuming $\beta$ independent of $k$.
   }
\end{table}

\subsection{DESI forecasts}

The DESI survey \cite{DESI:2016fyo,Vargas-Magana:2018rbb} is a ground telescope which will produce a spectroscopic map covering 14000 deg$^{2}$ of the sky.  For DESI we consider the LRG and ELG tracers, for which the main redshift coverage can be split into seven bins from $z=0.65$ to $z=1.25$ (central values), as detailed in Table~\ref{tab:survey-specs}. The number densities are simply the sum of the LRG and ELG forecasts, since in this present work we are not investigating the advantages of a multi-tracer approach.

\begin{table}
\centering
\small
\begin{tabular}{ccccccccc}
\hline
\multicolumn{9}{|c|}{{\small$k_{\mathrm{max}}=0.1\,h/$Mpc}}\tabularnewline
\hline
{\small$z$ } & {\small{}$\frac{\Delta E}{E}=\frac{\Delta\eta}{\eta}$ } & {\small$\Delta c_{0}$ } & {\small$\frac{\Delta b_{1}}{b_{1}}$ } & {\small$\Delta b_{2}$ } & {\small$\Delta b_{{\cal {G}}_{2}}$ } & {\small$\Delta b_{\Gamma_{3}}$} & $\frac{\Delta \sigma_f}{\sigma_f} $& $\Delta P_{\rm shot}$
\tabularnewline
\hline
\hline
 0.65 & 0.284 & 17500. & 9.27 & 705. & 323. & 916. & 1. & 1. \\
 0.75 & 0.246 & 15100. & 9.25 & 814. & 316. & 801. & 1. & 1. \\
 0.85 & 0.229 & 14000. & 9.13 & 853. & 322. & 802. & 1. & 1. \\
 0.95 & 0.215 & 14600. & 9.73 & 890. & 358. & 894. & 1. & 1. \\
 1.05 & 0.212 & 14500. & 10. & 1020. & 384. & 963. & 1. & 1. \\
 1.15 & 0.206 & 14700. & 10.8 & 1250. & 429. & 1000. & 1. & 1. \\
 1.25 & 0.199 & 14000. & 11.8 & 1330. & 446. & 965. & 1. & 1. \\
\tabularnewline
\end{tabular}

\begin{tabular}{ccccccccc}
\hline
\multicolumn{9}{|c|}{{\small$k_{\mathrm{max}}=0.2\,h/$Mpc}}\tabularnewline
\hline
{\small$z$ } & {\small{}$\frac{\Delta E}{E}=\frac{\Delta\eta}{\eta}$ } & {\small$\Delta c_{0}$ } & {\small$\frac{\Delta b_{1}}{b_{1}}$ } & {\small$\Delta b_{2}$ } & {\small$\Delta b_{{\cal {G}}_{2}}$ } & {\small$\Delta b_{\Gamma_{3}}$} & $\frac{\Delta \sigma_f}{\sigma_f} $ & $\Delta P_{\rm shot}$
\tabularnewline
\hline
\hline
 0.65 & 0.0845 & 345. & 1.15 & 14.6 & 7.5 & 23.8 & 0.663 & 1. \\
 0.75 & 0.0848 & 430. & 1.09 & 17. & 9.59 & 31.7 & 0.665 & 1. \\
 0.85 & 0.0893 & 577. & 1.21 & 21.7 & 13.4 & 47.1 & 0.746 & 1. \\
 0.95 & 0.0861 & 701. & 1.3 & 26.2 & 17.2 & 62.3 & 0.808 & 1. \\
 1.05 & 0.0849 & 828. & 1.47 & 32.2 & 21.9 & 78.6 & 0.867 & 0.999 \\
 1.15 & 0.0834 & 857. & 1.57 & 35.5 & 24.6 & 87.8 & 0.898 & 0.998 \\
 1.25 & 0.0839 & 849. & 1.68 & 38.2 & 26.4 & 95. & 0.914 & 0.997 \\
\tabularnewline
\end{tabular}

\begin{tabular}{ccccccccc}
\hline
\multicolumn{9}{|c|}{{\small$k_{\mathrm{max}}=0.3\,h/$Mpc}}\tabularnewline
\hline
{\small$z$ } & {\small{}$\frac{\Delta E}{E}=\frac{\Delta\eta}{\eta}$ } & {\small$\Delta c_{0}$ } & {\small$\frac{\Delta b_{1}}{b_{1}}$ } & {\small$\Delta b_{2}$ } & {\small$\Delta b_{{\cal {G}}_{2}}$ } & {\small$\Delta b_{\Gamma_{3}}$} & $\frac{\Delta \sigma_f}{\sigma_f} $ & $\Delta P_{\rm shot}$
\tabularnewline
\hline
\hline
  0.65 & 0.0138 & 19.9 & 0.148 & 0.63 & 0.508 & 0.967 & 0.101 & 0.96 \\
 0.75 & 0.0122 & 19.8 & 0.143 & 0.49 & 0.493 & 0.959 & 0.079 & 0.984 \\
 0.85 & 0.0162 & 26.8 & 0.172 & 0.821 & 0.683 & 1.65 & 0.1 & 0.971 \\
 0.95 & 0.0206 & 37.1 & 0.204 & 1.35 & 0.995 & 2.7 & 0.12 & 0.959 \\
 1.05 & 0.029 & 63.3 & 0.279 & 2.87 & 1.83 & 5.18 & 0.17 & 0.917 \\
 1.15 & 0.0349 & 94.4 & 0.344 & 4.57 & 2.91 & 8.39 & 0.203 & 0.9 \\
 1.25 & 0.0398 & 136. & 0.416 & 6.89 & 4.5 & 13.1 & 0.235 & 0.89 \\
\tabularnewline
\end{tabular}{\small{}\quad{}\caption{\label{tab:DESI-results}   Forecasts for the DESI (LRG+ELG) survey.
 }
}
\end{table}

\begin{figure}
    \centering
    \includegraphics[width=.63\columnwidth]{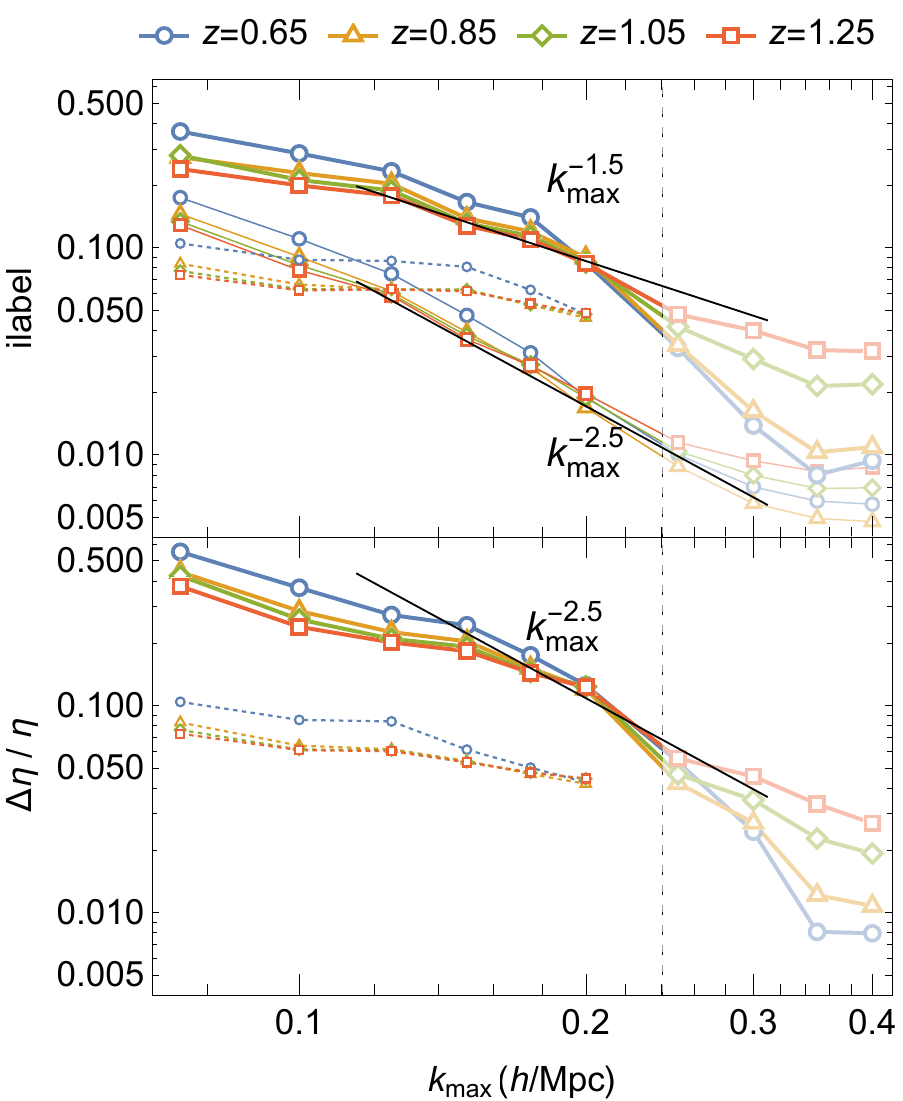}
    \caption{
    \emph{Top:} Fractional error in $\eta$ as a function of $k_{\rm max}$ for different redshifts for DESI. Thick lines are for the full one-loop non-linear power spectrum with free $\beta(k)$;  dotted lines for the naïve linear (tree-level) one; thin lines are for the $\beta$  $k$-independent case. The precision scaling goes like $k_{\rm max}^{-1.5}$ ($k_{\rm max}^{-2.5}$) for the free $\beta(k)$ (constant $\beta$) case. For $k_{\rm max} \gtrsim 0.25 h$/Mpc (see vertical dashed line) higher order corrections may be relevant.
    \emph{Bottom:} Same but for the case in which the counterterms $c_2$ and $\tilde{c}$ are added with uninformative priors and the FoG parameter $\sigma_f$ kept fixed to zero.         \label{fig:delta-eta-scaling}}
\end{figure}

In this and the next sections, all parameters, except  $\log\sigma_f$ and $P_{\rm shot}$,  are taken with non-informative priors (notice that when we take the logarithm of the  parameters, as per Eq.~\eqref{eq:parfull}, the prior refers to the logarithm). In practice, we assume a Gaussian prior with standard deviation much larger than the uncertainty in each parameter (relative for the log-parameters and absolute for the others) and check that the results do not change for wider priors.  For $\log\sigma_f$ and $P_{\rm shot}$  we adopt a Gaussian standard deviation of 1, the rationale being that these two parameters depend on physical properties that are, at least in principle, less related to the cosmological model and more under control. We explore some other choices in what follows.

The main DESI results are in Table~\ref{tab:DESI-results} (in this and all subsequent tables the quoted redshift is the  central one in the given bin). The key result is that one can estimate $\eta$ to a precision of around $8\div 9\%$ (for $k_{\rm max}=0.2\,h/$Mpc) and around $1\div 4\%$ (for $k_{\rm max}=0.3\,h/$Mpc) in all seven DESI redshift bins, regardless of the cosmological model. Figure~\ref{fig:delta-eta-scaling} illustrates how the DESI results on $\eta$ (and thus $E$) scale as a function of $k_{\rm max}$. We see that the constraints scale approximately as $k_{\rm max}^{-1.5}$ (which is intuitive as the number of modes increases as $k_{\rm max}^{3}$) and then flatten out at large $k_{\rm max}$ when the shot noise dominates.

We also compare in the same figure the linear case. Here and in the following, the linear case is obtained by fixing all the NL parameters to their fiducial (so it is not the same as just discarding the NL correction). In this linear case the trend saturates already at $k_{\rm max}\approx 0.1\,h$/Mpc. The main reason the NL case has a larger slope for $E(z)$ is that the bias and counterterm parameters all get better constrained very fast with increased $k_{\rm max}$. In fact, from Figures~\ref{fig:scaling-eta-others} and \ref{fig:scaling-eta-others-c2ctilde} in Appendix~\ref{app:scaling} we see that these nuisance parameters improve with powers typically between $-3$ and $-4$ of $k_{\rm max}$. This in turn  quickly suppresses the effects of the degeneracies (due to marginalization over them), and leads to a steeper slope for $E(z)$. In the linear case these parameters are all completely known from the onset, so a higher $k_{\rm max}$ does not lead to further degeneracy breaking.

Clearly, since in the linear case we fixed the NL  parameters, the constraints on $E(z)$ for the same $k_{\rm max}$ are much more stringent than in the full case in which the NL parameters are freely varied.
However, the constraints of the linear model  are   prior-dominated: relaxing the priors on the NL parameters, the bound on $\eta$ for $k_{\rm max}=0.1\,h/$Mpc worsens by around four- or five-fold  (we discuss the issue of prior choice below, see  Figure~\ref{fig:2D-contours}). So an important message of this paper is that neglecting the NL terms and  cutting at, say, $k_{\rm max}=0.1\,h/$Mpc is not a model-independent safe choice: the constraints can vary  substantially depending on the NL parameters.

Notice that the constraints  for $\sigma_f$ and $P_{\rm shot}$ are  prior-dominated for small $k_{\rm max}$. Decreasing the errors in both by a factor of 10 has only a small effect on the constraints on $E(z)$ for $k_{\rm max} = 0.2\,h$/Mpc, improving them by between 2\% and 9\% depending on the redshift. On the other hand, enlarging the errors by a factor of 10 increases the errors for the same $k_{\rm max}$ by between 28 and 44\%. Clearly, higher (lower) $k_{\rm max}$ decreases (increases) this prior susceptibility. To summarize, if a very loose 100\% prior can be justified in these parameters, this is already enough to extract most of the information on $E(z)$ with our method.

\begin{table}
\centering
\small
\begin{tabular}{ccccccccc}
\hline
\multicolumn{9}{|c|}{{\small$k_{\mathrm{max}}=0.1\,h/$Mpc}}\tabularnewline
\hline
{\small$z$ } & {\small{}$\frac{\Delta E}{E}=\frac{\Delta\eta}{\eta}$ } & {\small$\Delta c_{0}$ } & {\small$\frac{\Delta b_{1}}{b_{1}}$ } & {\small$\Delta b_{2}$ } & {\small$\Delta b_{{\cal {G}}_{2}}$ } & {\small$\Delta b_{\Gamma_{3}}$} & $\frac{\Delta \sigma_f}{\sigma_f} $ & $\Delta P_{\rm shot}$
\tabularnewline
\hline
\hline
1 & 0.131 & 6390. & 7.69 & 536. & 157. & 340. & 1. & 1. \\
 1.2 & 0.126 & 6150. & 8.96 & 661. & 178. & 332. & 1. & 1. \\
 1.4 & 0.121 & 5220. & 10.3 & 679. & 186. & 304. & 1. & 1. \\
 1.65 & 0.102 & 4830. & 11.8 & 712. & 221. & 311. & 1. & 0.999 \\\\
\tabularnewline
\end{tabular}

\begin{tabular}{ccccccccc}
\hline
\multicolumn{9}{|c|}{{\small$k_{\mathrm{max}}=0.2\,h/$Mpc}}\tabularnewline
\hline
{\small$z$ } & {\small{}$\frac{\Delta E}{E}=\frac{\Delta\eta}{\eta}$ } & {\small$\Delta c_{0}$ } & {\small$\frac{\Delta b_{1}}{b_{1}}$ } & {\small$\Delta b_{2}$ } & {\small$\Delta b_{{\cal {G}}_{2}}$ } & {\small$\Delta b_{\Gamma_{3}}$} & $\frac{\Delta \sigma_f}{\sigma_f} $ & $\Delta P_{\rm shot}$
\tabularnewline
\hline
\hline
 1 & 0.0569 & 405. & 1.09 & 15.7 & 11. & 43.7 & 0.785 & 0.993 \\
 1.2 & 0.0547 & 397. & 1.23 & 18.7 & 12.6 & 50.1 & 0.849 & 0.988 \\
 1.4 & 0.0558 & 418. & 1.51 & 25.6 & 15.8 & 60.5 & 0.891 & 0.98 \\
 1.65 & 0.0526 & 436. & 1.79 & 37.7 & 20.3 & 76. & 0.913 & 0.942 \\
\tabularnewline

\end{tabular}
\begin{tabular}{ccccccccc}
\hline
\multicolumn{9}{|c|}{{\small$k_{\mathrm{max}}=0.3\,h/$Mpc}}\tabularnewline
\hline
{\small$z$ } & {\small{}$\frac{\Delta E}{E}=\frac{\Delta\eta}{\eta}$ } & {\small$\Delta c_{0}$ } & {\small$\frac{\Delta b_{1}}{b_{1}}$ } & {\small$\Delta b_{2}$ } & {\small$\Delta b_{{\cal {G}}_{2}}$ } & {\small$\Delta b_{\Gamma_{3}}$} & $\frac{\Delta \sigma_f}{\sigma_f} $ & $\Delta P_{\rm shot}$
\tabularnewline
\hline
\hline
 1 & 0.0187 & 86.2 & 0.225 & 3.77 & 2.81 & 5.46 & 0.232 & 0.262 \\
 1.2 & 0.0256 & 167. & 0.345 & 8.25 & 6.33 & 13.4 & 0.345 & 0.268 \\
 1.4 & 0.0321 & 267. & 0.514 & 14.5 & 11.4 & 27.4 & 0.427 & 0.209 \\
 1.65 & 0.0331 & 364. & 0.724 & 22.5 & 17.7 & 49.4 & 0.536 & 0.122 \\
\tabularnewline

\end{tabular}{\small{}\quad{}\caption{\label{tab:Euclid-results}   Forecasts for the Euclid survey.
 }
}
\end{table}

\subsection{Euclid forecasts}

We move now to forecasts for a Euclid-like galaxy survey. The Euclid survey is a space telescope, to be launched in 2023,  that will map 15000 deg$^{2}$ of the sky  \cite{laureijs2011euclid}. The redshift bins we employ here and their main properties are in Table \ref{tab:survey-specs} (see \cite{2020A&A...642A.191E}). The results for Euclid are in Table~\ref{tab:Euclid-results}. One can estimate $\eta$ to a precision of around $5-6\%$ (for $k_{\rm max}=0.2\,h/$Mpc) and around $2-3\%$ (for $k_{\rm max}=0.3\,h/$Mpc) in all four Euclid redshift bins, regardless of the cosmological model.

\begin{figure}
    \centering
    \includegraphics[width=.63\columnwidth]{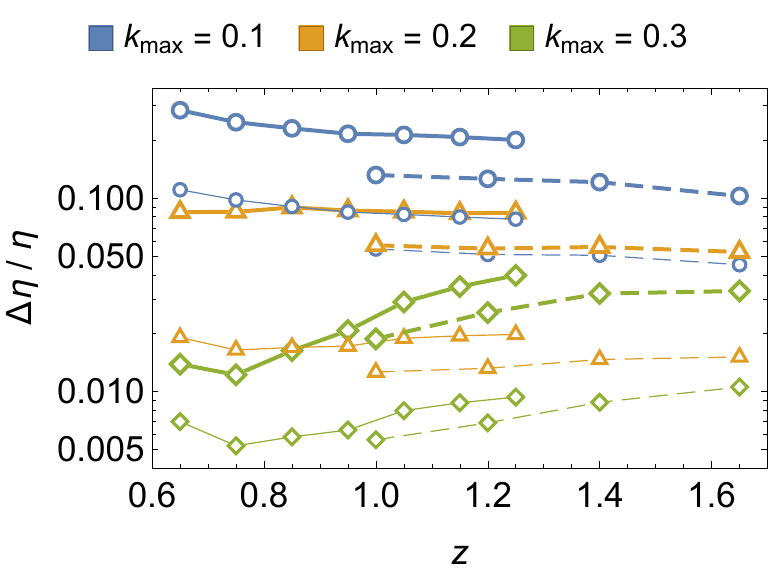}
    \caption{
    Fractional error in $\eta$ as a function of redshift for different values of $k_{\rm max}$ (in units of $h$/Mpc) and assuming either $\beta$ a free function of $k$ (thick lines) or $\beta$  independent of $k$ (thin lines). Solid (dashed) lines represent the DESI (Euclid) forecasts.
    \label{fig:delta-log-eta}}
\end{figure}

The results for both DESI and Euclid as a function of redshift are summarized in Figure~\ref{fig:delta-log-eta}. As anticipated, the gain from $k_{\rm max}=0.1\,h/$Mpc to $k_{\rm max}=0.2\,h/$Mpc is around $2\div 3$ across the entire $z$ range. The gain increases by a total factor from 4 to 9 when cutting off at  $k_{\rm max}=0.3\,h/$Mpc.  We stress again that these gains are achieved for uninformative  priors on all bias and counterterm parameters.

\subsection{Constraints on the growth rate}

\begin{figure}
    \centering
    \includegraphics[width=.63\columnwidth]{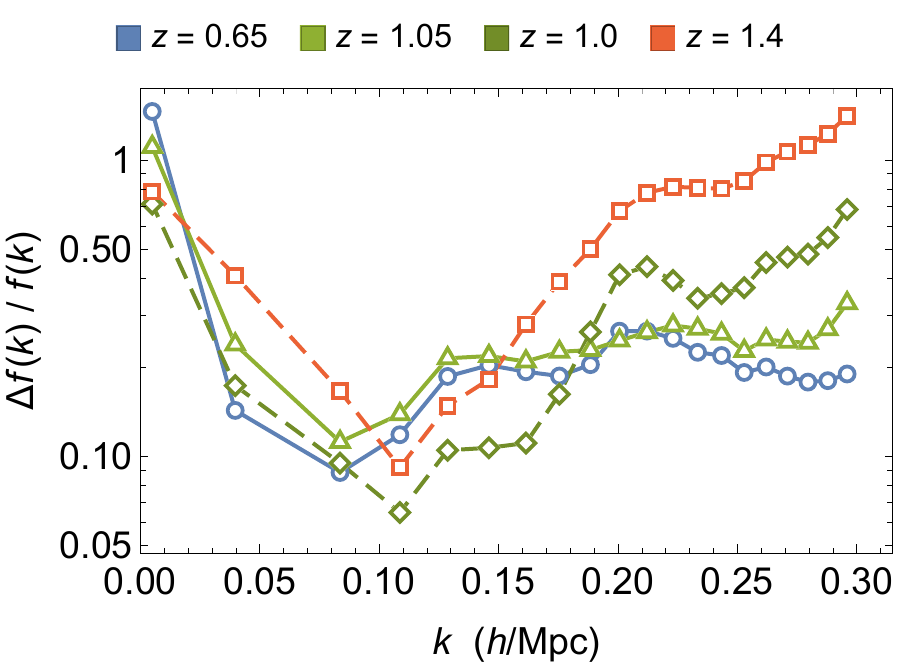}
    \caption{
    Fractional error forecasts with DESI (solid) Euclid (dashed lines) for $f(k)$ as a function of redshift for different values of $k$ when assuming $\beta$ a free function of $k$ and $k_{\rm max} = 0.3 h$/Mpc. A lower $k_{\rm max} = 0.2 h$/Mpc increases errors by factors of $\sim 3$.
    \label{fig:delta-logfofk}}
\end{figure}

Once we have the Fisher matrix marginalized over all parameters except $\beta(k)$ and $b_1$, we can extract the errors on $f(k)$ by transforming from $X=\{\beta_1=f_1/b_1,\beta_2=f_2/b_1,...b_1\}$ to $Y=\{f_1,f_2,..,b_1\}$, where $\beta_i,f_i$ are the values at each $k$ bin. Since our parameters are actually the $\log$ of these variables, the Jacobian of the transformation is
\begin{equation}
    J_{ij}=\frac{Y_j}{X_i}\frac{\partial X_i}{\partial  Y_j}\,,
\end{equation}
and the projected matrix is
\begin{equation}
    F_Y = J^T F_X J \,.
\end{equation}
We find however that the relative constraints on $\beta_i$, and consequently on $f_i$, are weak, reaching less than 10$\%$ in some $k$-bins only if $k_{\rm max}=0.3\,h/$Mpc (see Figure~\ref{fig:delta-logfofk}). Combining several $k-$bins the errors would of course reduce.

One obvious way to improve the constraints is assuming the simplest, but still physically interesting, case, namely taking $\beta$ (and therefore $f$) to depend only on $z$ and not on $k$. This is in fact a good approximation in  models that do not depart too much from $\Lambda$CDM, as e.g. uncoupled dynamical dark energy, or in modifications of General Relativity such as the nDGP model.  The constraints on $\beta$ improve very significantly, as shown in Table~\ref{tab:Euclid-betaconst-full}. In particular, the uncertainty on $E$ decreases by three to four  times at $k_{\rm max}=0.2\,h/$Mpc, reaching 1--2\%.

\begin{figure}
    \centering
    \includegraphics[width=.63\columnwidth]{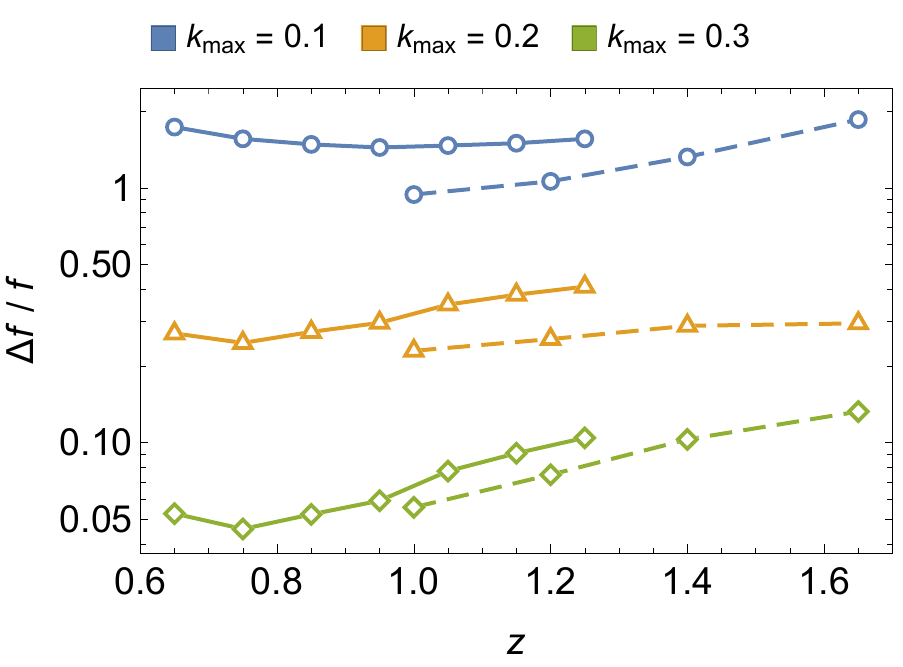}
    \caption{
    Fractional error in $f$ as a function of redshift for different values of $k_{\rm max}$ and assuming  $\beta$ independent of $k$. Solid (dashed) lines represent the DESI (Euclid) forecasts.
    \label{fig:delta-logf}}
\end{figure}

The constraints on $f$ improve as well but  remain still unsatisfactory unless high $k$ bins remain accurate at one-loop: we find $\Delta f/f\approx 0.2\div 0.5$ for $k_{\rm max}=0.2\,h/$Mpc and $\Delta f/f\approx 0.05\div 0.2$ for $k_{\rm max}=0.3\,h/$Mpc (see Figure~\ref{fig:delta-logf}).

\begin{table}
\small
\centering
\begin{tabular}{cccccccccc}
\hline
\multicolumn{10}{|c|}{$k_{\mathrm{max}}=0.2\,h/$Mpc}\tabularnewline
\hline
$z$ & $\frac{\Delta \beta }{\beta }$& $\frac{\Delta E}{E}=$  $\frac{\Delta \eta}{\eta}$ & $\Delta c_{0}$ & $\frac{\Delta b_{1}}{b_{1}}$
& $\Delta b_{2}$
& $\Delta b_{{\cal{G}}_2}$
& $\Delta b_{\Gamma_3}$ & $\frac{\Delta \sigma_f}{\sigma_f} $ & $\Delta P_{\rm shot}$
\tabularnewline
\hline
\multicolumn{10}{c}{DESI}\tabularnewline
\hline
 0.65 & 0.049 & 0.019 & 37.7 & 0.264 & 1.41 & 1.62 & 3.55 & 0.584 & 0.977 \\
 0.75 & 0.0448 & 0.0164 & 41.8 & 0.242 & 1.59 & 1.82 & 3.59 & 0.533 & 0.994 \\
 0.85 & 0.0492 & 0.0169 & 54.2 & 0.268 & 2.12 & 2.35 & 4.73 & 0.551 & 0.984 \\
 0.95 & 0.0532 & 0.0172 & 68.8 & 0.291 & 2.81 & 2.97 & 6.13 & 0.552 & 0.971 \\
 1.05 & 0.0629 & 0.0188 & 92.8 & 0.343 & 3.99 & 3.98 & 8.68 & 0.575 & 0.914 \\
 1.15 & 0.0691 & 0.0194 & 116. & 0.375 & 5.32 & 4.98 & 11.2 & 0.558 & 0.876 \\
 1.25 & 0.0748 & 0.0197 & 141. & 0.403 & 6.93 & 6.07 & 14.1 & 0.529 & 0.846 \\
\hline
\multicolumn{10}{c}{Euclid}\tabularnewline
\hline
 1 & 0.0444 & 0.0126 & 59.5 & 0.225 & 2.23 & 2.24 & 4.69 & 0.374 & 0.741 \\
 1.2 & 0.0513 & 0.0132 & 87. & 0.25 & 3.77 & 3.33 & 7.68 & 0.339 & 0.621 \\
 1.4 & 0.0625 & 0.0146 & 114. & 0.281 & 6.09 & 4.55 & 11.7 & 0.365 & 0.496 \\
 1.65 & 0.0739 & 0.0151 & 121. & 0.285 & 8.87 & 5.32 & 15.1 & 0.434 & 0.307 \\
\hline
\end{tabular}
\caption{\label{tab:Euclid-betaconst-full}   Results for DESI  and Euclid  with $\beta$ independent of $k$.   }
\end{table}

\subsection{On the robustness of the results}

We carried out a number of tests of the robustness of our results, focusing on how the constraints on $E$ vary from case to case in the Euclid-like survey, and  adopting the $k_{\rm max}=0.2\,h/$Mpc reference.

First, we evaluated the constraints for a $\Lambda$CDM spectrum without wiggles, obtained with the Eisenstein-Hu fitting formula of Ref.~\cite{Eisenstein:1997ik}. We find that the $\Delta E/E$ results degrade by 2\% to 13$\%$ depending on the redshift bin, with an average of 8\%,  (as can be seen by comparing the middle panel in Table~\ref{tab:Euclid-results}  with Table \ref{tab:Euclid-nowiggle}). This mild dependence is to be expected since  our method is weakly sensitive to the exact location of the BAO wiggles, because the (binned) spectrum shape is marginalized over. It is sensitive however to the spectrum slope (since the AP effect depends on the spectrum slope) and therefore also to the height of the wiggles (see Appendix E for more on this point). We conclude that the degradation if we included BAO damping by large scale bulk motions on the wiggly PS would be likely less than the average 8$\%$ we find for a completely de-wiggled PS.

Second, we tested discarding the off-diagonal derivatives with respect to $P$ (i.e. we put $\partial P_{gg}(k)/\partial P(q)\propto \delta_{pq}$), finding variations  below 3$\%$. Third, we increased the number of $k$-bins from 20 to 40 and  we found very small differences, less than $4\%$, (compare Table \ref{tab:Euclid-40bins} with the central panel of Table \ref{tab:Euclid-results}). This  shows that the results are not influenced by the $k$-bin size or by their smoothing of the wiggle features (see Appendix E).

Fourth, we adopted different fiducials for the NL terms (except $b_1$),
namely the same BOSS fiducials but randomly displaced by $0.5\,\sigma$ in each parameter, see Table \ref{tab:Euclid-alt-fid2}. The constraints on $E$ vary by 5$\%$ on average.

Fifth, we find that relaxing the prior on $P_{\rm shot}$ to an uninformative one, the constraints on $E$ worsen by 15 to 50\%, while $\Delta P_{\rm shot}$ becomes roughly 8 for the $z=1$  and  3 for $z=1.65$ Euclid bins.

Sixth, we adopted the alternative UV parametrization in terms of $c_2,\tilde c$  instead of the FoG damping -- see Eq.~\eqref{eq:UV-2}. The results are in Table \ref{tab:Euclid-results-c2c4} (and in Figure~\ref{fig:delta-eta-scaling}, bottom panel, for DESI), assuming again uninformative priors except for $P_{\rm shot}$. This model trades-off $\sigma_f$ for two extra unconstrained parameters, so it has one extra parameter with respect to the exponential FoG case and the bounds on $E$ are consequently quite weaker,  to wit by on average 40\% (20\%) for $k_{\rm max}=0.2 \,h$/Mpc ($0.3 \,h$/Mpc).

\begin{table}
    \centering
    \begin{tabular}{ccccccccc}
    \hline
    \multicolumn{9}{|c|}{{\small$k_{\mathrm{max}}=0.2\,h/$Mpc}}\tabularnewline
    \hline
    {\small$z$ } & {\small{}$\frac{\Delta E}{E}=\frac{\Delta\eta}{\eta}$ } & {\small$\Delta c_{0}$ } & {\small$\frac{\Delta b_{1}}{b_{1}}$ } & {\small$\Delta b_{2}$ } & {\small$\Delta b_{{\cal {G}}_{2}}$ } & {\small$\Delta b_{\Gamma_{3}}$} & $\frac{\Delta \sigma_f}{\sigma_f} $ & $\Delta P_{\rm shot}$
    \tabularnewline
    \hline
    \hline
      1 & 0.0642 & 1120. & 1.15 & 37.5 & 29.1 & 81.4 & 0.776 & 0.993 \\
     1.2 & 0.0603 & 1220. & 1.56 & 52. & 37.5 & 95.8 & 0.825 & 0.99 \\
     1.4 & 0.0588 & 1540. & 2.13 & 82. & 55.8 & 131. & 0.878 & 0.984 \\
     1.65 & 0.0538 & 1690. & 2.61 & 114. & 73.7 & 161. & 0.916 & 0.959 \\
    \hline
    \end{tabular}
    \caption{\label{tab:Euclid-nowiggle}   Representative results for the Euclid survey  assuming a no-wiggle spectrum.}
\end{table}

\begin{table}
    \centering
    \begin{tabular}{ccccccccc}
    \hline
    \multicolumn{9}{|c|}{{\small$k_{\mathrm{max}}=0.2\,h/$Mpc}}\tabularnewline
    \hline
    {\small$z$ } & {\small{}$\frac{\Delta E}{E}=\frac{\Delta\eta}{\eta}$ } & {\small$\Delta c_{0}$ } & {\small$\frac{\Delta b_{1}}{b_{1}}$ } & {\small$\Delta b_{2}$ } & {\small$\Delta b_{{\cal {G}}_{2}}$ } & {\small$\Delta b_{\Gamma_{3}}$} & $\frac{\Delta \sigma_f}{\sigma_f} $ & $\Delta P_{\rm shot}$
    \tabularnewline
    \hline
    \hline
       1 & 0.0572 & 401. & 1.14 & 15.5 & 10.8 & 44.1 & 0.79 & 0.993 \\
     1.2 & 0.0559 & 395. & 1.29 & 18.7 & 12.6 & 50.6 & 0.855 & 0.988 \\
     1.4 & 0.0573 & 416. & 1.57 & 25.4 & 15.8 & 61.3 & 0.895 & 0.981 \\
     1.65 & 0.0544 & 434. & 1.86 & 37.1 & 20.3 & 77.2 & 0.915 & 0.944 \\
    \hline
    \end{tabular}
    \caption{\label{tab:Euclid-40bins}   Representative results for the Euclid survey  with 40 $k$-bins instead of 20.}
\end{table}

\begin{table}
    \centering
    \begin{tabular}{ccccccccc}
    \hline
    \multicolumn{9}{|c|}{{\small$k_{\mathrm{max}}=0.2\,h/$Mpc}}\tabularnewline
    \hline
    {\small$z$ } & {\small{}$\frac{\Delta E}{E}=\frac{\Delta\eta}{\eta}$ } & {\small$\Delta c_{0}$ } & {\small$\frac{\Delta b_{1}}{b_{1}}$ } & {\small$\Delta b_{2}$ } & {\small$\Delta b_{{\cal {G}}_{2}}$ } & {\small$\Delta b_{\Gamma_{3}}$} & $\Delta \sigma_f$ & $\Delta P_{\rm shot}$
    \tabularnewline
    \hline
    \hline
     1 & 0.052 & 433. & 1.24 & 15.5 & 9.87 & 39.8 & 0.594 & 0.989 \\
     1.2 & 0.0522 & 553. & 1.57 & 20.7 & 14.1 & 60.8 & 0.78 & 0.987 \\
     1.4 & 0.0532 & 543. & 1.96 & 24.6 & 15.8 & 74.2 & 0.86 & 0.979 \\
     1.65 & 0.0514 & 541. & 2.39 & 35.5 & 19. & 90.8 & 0.885 & 0.939 \\
    \hline
    \end{tabular}
    \caption{\label{tab:Euclid-alt-fid2}    Representative results for Euclid survey with alternative fiducials for all NL parameters except $b_1$, each chosen 0.5$\sigma$ away from the BOSS fiducial values. }
\end{table}

\begin{table}[t]
    \small
    \centering
    \begin{tabular}{cccccccccc}
    \hline
    \multicolumn{10}{|c|}{$k_{\mathrm{max}}=0.2\,h/$Mpc}\tabularnewline
    \hline
    $z$ &$\frac{\Delta E}{E}=\frac{\Delta \eta}{\eta}$ & $\Delta c_{0}$ & $\Delta c_{2}$& $\Delta \tilde{c}$ & $\frac{\Delta b_{1}}{b_{1}}$
    & $\Delta b_{2}$
    & $\Delta b_{{\cal{G}}_2}$
    & $\Delta b_{\Gamma_3}$ & $\Delta P_{\rm shot}$
    \tabularnewline
    \hline
    \hline
      1 & 0.0776 & 874. & 218. & 2490. & 1.83 & 31.9 & 8.31 & 79.8 & 0.996 \\
     1.2 & 0.0771 & 1110. & 303. & 2950. & 2.22 & 46. & 9.62 & 114. & \
    0.992 \\
     1.4 & 0.0791 & 1420. & 402. & 3440. & 2.72 & 67.5 & 12.7 & 162. & \
    0.987 \\
     1.65 & 0.0724 & 1680. & 493. & 3770. & 3.14 & 93.1 & 17.4 & 227. & \
    0.962 \\
    \hline
    \end{tabular}
    \quad
    \quad
    \caption{\label{tab:Euclid-results-c2c4}   Euclid forecasts for the case in which the FoG are modelled with $c_2,\tilde{c}$.}
\end{table}

\begin{figure}
    \centering
    \includegraphics[width=1.0\columnwidth]{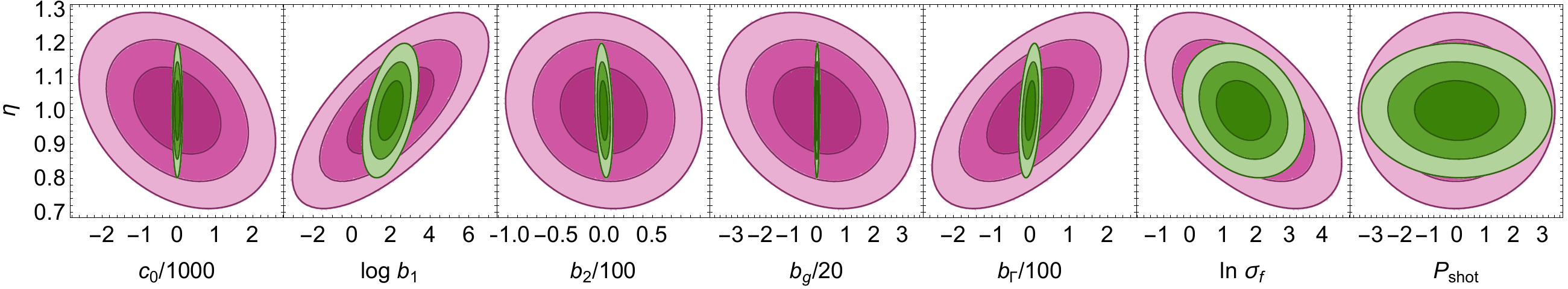}
    \caption{
    1, 2 and 3$\sigma$ confidence levels for $\eta$ and other variables for the case of DESI at $z=1.05$ and with $k_{\rm max} = 0.2\,h$/Mpc. The magenta contours are assuming completely uninformative priors for the bias and counterterm parameters, whereas the green contours assume priors with uncertainties in each variable exactly like their fiducial values, except for $b_\Gamma$, for which we assume $\sigma b_\Gamma = 10$. In all cases we keep to the default priors for the FoG parameters $\sigma_f$ and $P_{\rm shot}$.
    \label{fig:2D-contours}}
\end{figure}

Figure~\ref{fig:2D-contours} illustrates the dependency of the constraints on the choice of priors, using DESI at $z=1.05$ with $k_{\rm max} = 0.2 h/$Mpc as an example. Priors are assumed to be independent and Gaussian (as per the Fisher Matrix treatment) in the parameters $\{c_0,\,\log b_1,\,b_2,$ $\,b_{{\cal G}_2},\,b_{\Gamma_3}, \log \sigma_f, \,P_{\rm shot}\}$. The magenta contours assume completely uninformative priors for the bias and counterterm parameters, where as the green contours assume priors with uncertainties in each variable exactly like their fiducial values, except for $b_\Gamma$, for which we assume $\Delta b_\Gamma = 10$. In all cases we keep to the default priors for the FoG parameters $\sigma_f$ and $P_{\rm shot}$. Although the choice above for the tight priors is completely arbitrary, it just serves as an illustrative example. The important point is that while the much tighter priors result in much smaller errors on the bias and counterterm parameters, the constraints on $\eta$ are very robust, with very little dependency on these priors. The change in $\Delta \eta$ is only around $30\%$ when going from the extreme limits of completely uninformative to Delta Dirac priors on the bias and counterterms. This is in line with previous findings in~\cite{Ivanov:2019pdj}, according to which the priors on the bias parameters have little effect on the cosmological ones.

So far we have not addressed the issue of the correlations among the cosmological and nuisance parameters. They are discussed in Appendix~\ref{app:corr}.

\subsection{Constraints on $P(k)$ and $\beta(k)$}

\begin{figure}
    \centering
    \includegraphics[width=.48\columnwidth]{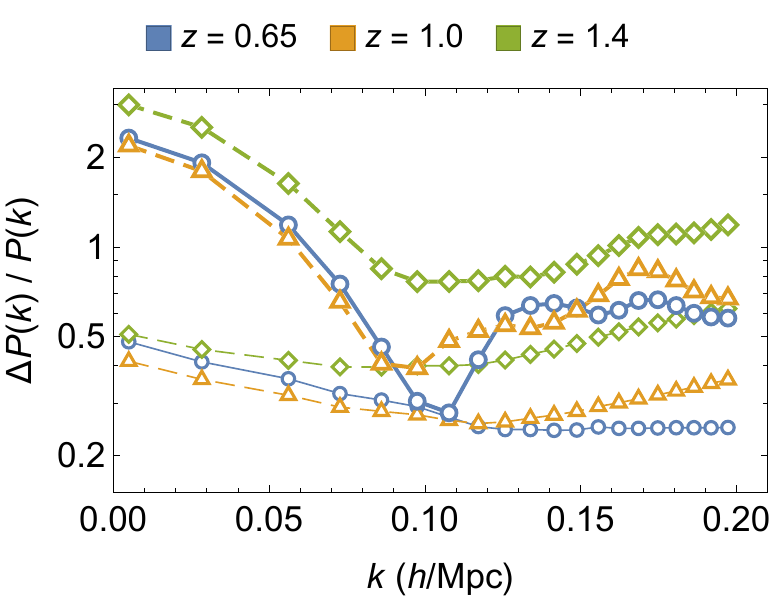}\quad
    \includegraphics[width=.48\columnwidth]{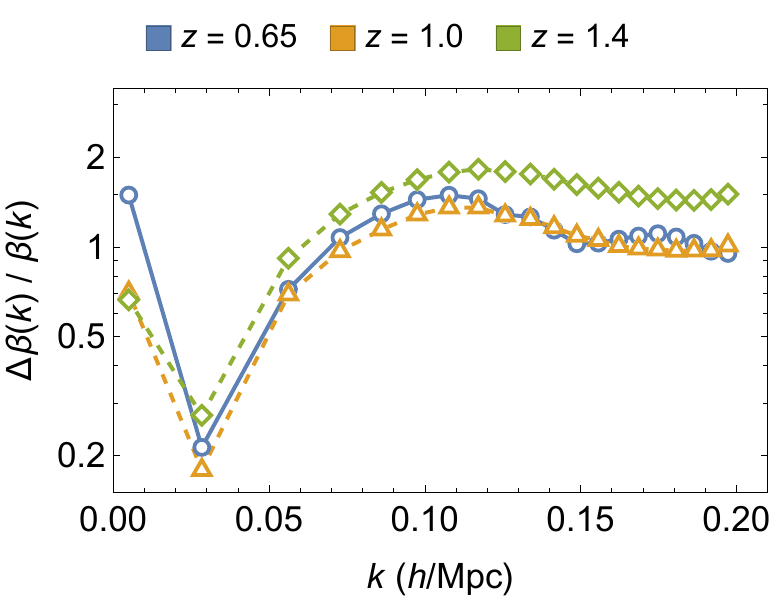}
    \caption{
    \emph{Left:} Fractional error in $P(k)$ as a function of $k$ for different redshifts, for $k_{\rm max}~=~0.2\,h$/Mpc. Solid (dashed) lines represent the DESI (Euclid) forecasts. Thick (thin) lines      assume $\beta$ a free function of $k$ ($\beta$  independent of $k$).   \emph{Right:} same for $\beta(k)$. As can be seen, in the free $\beta(k)$ case the individual S/N in each bin is usually only ${\cal O}(1)$. For $k_{\rm max} = 0.3\,h$/Mpc the uncertainties, in the same $k$ and $z$ range, shrink on average by 3.8.
    \label{fig:delta-logP-logB}}
\end{figure}

Finally, in Figure~\ref{fig:delta-logP-logB}, we display the expected uncertainties on the matter linear power spectrum $P(k)$ and the redshift distortion $\beta(k)$ in various $k$-bins for $k_{\rm max}=0.2\,h/$Mpc. As it can be seen, these errors are very large, unless one assumes a scale independent $\beta$, which is depicted with thin lines. These uncertainties can be reduced further by going to higher $k_{\rm max}$ or by operating with larger $k$-bins. Since the emphasis of this paper is on $H/H_0$ and $f$, however, we will not investigate further these issues here. We remark however that our method is still able to produce  precision measurements of $H/H_0$ even if the individual errors in $P(k)$ and $\beta(k)$ remain large.

\section{Conclusions}

Present and future large-scale cosmological data are opening a new era of precision while at the same time showing how difficult is to improve upon accuracy, given not just the observational systematics but also the many uncertainties about theoretical modelling. In this work we tried to work in both directions: increase the precision by including mildly non-linear scales, and increase the accuracy by remaining agnostic with respect to the cosmological model. These goals have been  implemented through a novel non-linear and model-independent Fisher matrix approach, in which the parameters are the linear power spectrum and the growth rate in space and time bins, along with a host of bias and counterterm parameters, and the Alcock-Paczyński parameter $\eta$ proportional to the combination $H(z)D(z)$. Given an external determination of the dimensionless distance $H_0 D$, we can produce model-independent estimates of  $E(z)\equiv H(z)/H_0$ and $f$. Crucially, the latter quantity can be estimated only by going beyond linearity.

We first compared our method to the real-data analysis of BOSS, and then to the  Euclid and DESI future surveys, covering the redshift range 0.6-1.8. We obtained several results, that we now summarize.
\begin{itemize}
    \item We find that adopting only a linear spectrum with a cut at, say, $k_{\rm max}=0.1\,h/$Mpc, is not a model-independent choice and risks a significant bias in the results. Depending on assumptions on the prior for the non-linear terms, the constraints on $E$ can vary by roughly a factor of four.
    \item Adopting the NL corrections, we find a large improvement in the model-independent estimation of $E$, which can go from a factor of a few to nine when moving from $k_{\rm max}=0.1\,h/$Mpc to  $k_{\rm max}=0.3\,h/$Mpc.
    \item With $k_{\rm max}=0.2\,h/$Mpc, the uncertainty on $E$ turns out to be around 5--9\% in each $z$-bin; with  $k_{\rm max}=0.3\,h/$Mpc one reaches 1--4\% precision.   Note that the knowledge of both $H(z)$ and $D(z)$ allows a measurement of the present space curvature $\Omega_{k0}$.
    \item The growth rate $f(k)$ is difficult to measure precisely if it is allowed to be freely $k$-dependent, with typical uncertainties between $10$ and $50\%$  for each of 20 $k$-bins even in the case of $k_{\rm max}=0.3\,h/$Mpc, and often much larger; if we instead restrict the analysis to the $k$-independent case, we find  errors of 20--40\% (5--13\%) for both Euclid and DESI if $k_{\rm max}=0.2\,h/$Mpc ($0.3\,h/$Mpc).
\end{itemize}
We remark again that these results do not depend on a cosmological model, nor on parametrized bias functions, nor on priors on the NL parameters. The constraints come essentially from the AP effect and are based therefore, ultimately, on the assumption of statistical isotropy.

Several improvements and extensions can be devised. For instance, we should  assess up to which scale is  the NL correction acceptable when assuming a model independent approach making use of $N$-body simulations for different cosmologies. Further, we can adapt this method to the parameters of standard $\Lambda$CDM and its popular variants, for which we expect similar gains. We can also extend the formalism to include multi-tracing techniques, or  velocity fields (see for instance~\cite{Qin:2019axr}), or a combination of both as recently advocated in~\cite{Quartin:2021dmr,Alfradique:2022tox}. We plan to address these questions in future work.

\section*{Acknowledgments}

LA acknowledges support from DFG project  456622116. MQ is supported by the Brazilian research agencies FAPERJ, CNPq (Conselho Nacional de Desenvolvimento Científico e Tecnológico) and CAPES. This study was financed in part by the Coordenação de Aperfeiçoamento de Pessoal de Nível Superior - Brasil (CAPES) - Finance Code 001. We acknowledge support from the CAPES-DAAD bilateral project  ``Data Analysis and Model Testing in the Era of Precision Cosmology''. Many integrals have been performed using the CUBA routines \cite{Hahn_2005} by T. Hahn (http://feynarts.de/cuba).

\appendix

\section{Non-linear $P_{gg}$}\label{app:nonlin-P}

In this appendix we specify the model for the galaxy power spectrum used in our analysis, Eq.~(\ref{Pgg}). The standard perturbation theory one-loop corrections are
\begin{align}
    P^{\rm 1loop}(k,\mu, z) &=P_{13}(k,\mu, z)+ P_{22}(k,\mu, z)\,,
\end{align}
Where, omitting the redshift dependence,
\begin{align}
P_{13}(k,\mu) &\equiv
6 \,Z_1({\bf k})P(k) \int_{\bf q} Z_{3}(\mathbf{k},\mathbf{q},-\mathbf{q})\, P(q)\,,\\
P_{22}(k,\mu) &\equiv
2 \int_{\bf q} Z_{2}^{2}(\mathbf{q},\mathbf{k}-\mathbf{q}) \,P(q)P(|\mathbf{k}-\mathbf{q}|)\,,
\label{P22}
\end{align}
and where we have defined
\begin{equation}
 \int_{\bf q}\equiv\int\frac{d^3 q}{(2\pi)^3}\,.
\end{equation}
The biased and RSD-corrected kernels are (see e.g.~\cite{Ivanov:2019pdj})
\begin{align}
    Z_1({\bf k}) =&\left(1+ \mu^2 \beta(k)\right) b_1\,,
    \label{z1}\\
    Z_{2}(\mathbf{k}_1,\mathbf{k}_2)  = \,&b_1 \bigg\{ F_{2}(\mathbf{k}_{1},\mathbf{k}_{2})+\beta(k)\mu^{2}G_{2}(\mathbf{k}_{1},\mathbf{k}_{2})\nonumber \\
    & +b_1 \frac{\beta(k) \mu k}{2}\left[\frac{\mu_{1}}{k_{1}}\big(1+\beta(k_{2})\mu_{2}^{2}\big)+\frac{\mu_{2}}{k_{2}}\big(1+\beta(k_{1})\mu_{1}^{2}\big)\right]\bigg\}+\frac{b_{2}}{2}+b_{{\cal G}_2}S_{1}(\mathbf{k}_{1},\mathbf{k}_{2})\,,
\end{align}
(already symmetrized) and
\begin{align}
    Z_{3}(\mathbf{k}_{1},\mathbf{k}_{2},\mathbf{k}_{3})
    \!= \, & b_1\bigg\{ F_{3}(\mathbf{k}_{1},\mathbf{k}_{2},\mathbf{k}_{3})+\beta(k) \mu^{2}G_{3}(\mathbf{k}_{1},\mathbf{k}_{2},\mathbf{k}_{3})\nonumber\\
    &\;+\! b_1 \beta(k) \mu k \frac{\mu_3}{k_3}\Big[ F_{2}(\mathbf{k}_{1},\mathbf{k}_{2})+\beta(k_{12})\mu_{12}^{2}G_{2}(\mathbf{k}_{1},\mathbf{k}_{2})\Big]\nonumber \\
    &\;+\! b_1 \beta(k) \mu k \big(1+\beta(k_1) \mu_{1}^{2}\big)\frac{\mu_{23}}{k_{23}}G_{2}(\mathbf{k}_{2},\mathbf{k}_{3}) \!+\! b_1^2 \frac{\big[\beta(k) \mu k\big]^{2}}{2}\big(1+\beta(k_1) \mu_{1}^{2}\big)\frac{\mu_{2}}{k_{2}}\frac{\mu_{3}}{k_{3}}\bigg\}\nonumber \\
    & \!+2b_{{\cal G}_2}S_{1}(\mathbf{k}_{1},\mathbf{k}_{2}+\mathbf{k}_{3})F_{2}(\mathbf{k}_{2},\mathbf{k}_{3})+b_1 b_{{\cal G}_2}\beta(k)\mu k\frac{\mu_{1}}{k_{1}}S_{1}(\mathbf{k}_{2},\mathbf{k}_{3})\nonumber \\
    & \!+2b_{\Gamma_3}S_{1}(\mathbf{k}_{1},\mathbf{k}_{2}+\mathbf{k}_{3})(F_{2}(\mathbf{k}_{2},\mathbf{k}_{3})-G_{2}\big(\mathbf{k}_{2},\mathbf{k}_{3})\big)\,,
 \label{z3}
\end{align}
(to be symmetrized), where ${\bf k} ={\bf k}_1+{\bf k}_2$ in $Z_2$ and   ${\bf k} ={\bf k}_1+{\bf k}_2+{\bf k}_3 $ in $Z_3$, $\mu_i\equiv \mu({\bf k}_i)$ ($i=1,\dots, 3$) are defined as below Eq.~(\ref{Pgg}), $k_{12}=|{\bf k}_1+{\bf k}_2|$, $\mu_{12}\equiv \mu({\bf k}_{12})$, and so on.  $F_{2,3}$ and $G_{2,3}$ are the density and velocity kernels at second and third order, respectively, in standard perturbation theory.

We will employ the kernels for the Einstein de Sitter (EdS) cosmology, which can be found for instance in \cite{Bernardeau_2002}. Indeed, it is well known that the time dependence of the 1-loop terms is reproduced at the subpercent level by using EdS kernels and the correct growth factors (see for instance \cite{Pietroni08} for $\Lambda$CDM and  \cite{Bose:2018orj} for some modified gravity models). Moreover, it is reasonable to expect that our model independent approach should be largely insensitive to the tiny residual cosmology dependent effects not captured by the EdS approximation for the shape of the perturbative kernels. Terms in $b_{2}$ in Eq.~(\ref{z3}) have been discarded because they contribute to the power spectrum with terms degenerate with others. We also ``renormalized'' the $P_{22}$ integral by subtracting a constant $P_{22}(k\to 0)$ induced by the $b_2$ term.

Finally, we defined
\begin{align}
    S_{1}(\mathbf{k}_{1},\mathbf{k}_{2})&=\frac{(\mathbf{k}_{1}\cdot\mathbf{k}_{2})^{2}}{k_{1}^{2}k_{2}^{2}}-1\,.
\end{align}
Then we also add the  UV counterterms as in Ref. \cite{Ivanov:2019pdj}, adapted to our notation
\begin{align}
    P^{\rm UV}(k,\mu,z)&
    =-2P(k)(c_{0}^{2}k^{2}+\frac{3}{2}c_{2}^{2}\mu^{2}k^{2}-\frac{1}{2}\tilde{c}\, b_{1}^{6}\beta^{4}\mu^{4}k^{4}(1+\beta\mu^{2})^{2})
    \,.\label{eq:UV-2}
\end{align}

\section{Expressions for the derivatives \label{app:der}}

\subsection{Derivatives with respect to $k,\mu$}

Every $k,\mu$ depend on $\log \eta(z)$ through the AP effect.
On the fiducial, we have
\begin{align}
 \frac{\partial P_{gg}}{\partial \log \eta} & \,=\,\frac{\partial P_{gg}}{\partial k}  \frac{\partial k}{\partial \log \eta}+\frac{\partial P_{gg}}{\partial\mu}\frac{\partial\mu}{\partial \log \eta}\,,\nonumber \\
 & \,=\,\frac{\partial P_{gg}}{\partial k}k\mu^{2}+\frac{\partial P_{gg}}{\partial\mu}\mu(1-\mu^{2})\,.
\end{align}

\subsection{Derivatives with respect to $P$}

We need to take derivatives of $P_{gg}(k,\mu,z)$  with respect to
$P(k)$. We first consider $P^{\rm lin}(k,\mu,z)$, see eqs.~(\ref{Pgg}) and (\ref{z1}). We need to take the functional derivative
\begin{equation}
   \Delta q \frac{\partial P(k)}{\partial P(q)} = \Delta q\, \delta_D(k-q)\,,
\end{equation}
where we have multiplied by the bin width $\Delta q$ in order to work with dimensionless quantities. When considering discrete bins, one has
\begin{equation}
    \Delta q \,\delta_D(k-q) \to \delta_{kq}\,.
\end{equation}
Therefore we have
\begin{equation}
    \Delta q\frac{\partial P^{\rm lin}({\bf k})}{\partial P(q)}= Z_1({\bf k})^2 \Delta q \,\delta_D(k-q) \,.
\end{equation}
Then we consider  $P_{22}$, defined in Eq.~(\ref{P22}).
Using the symmetry of the integral with respect to ${\bf k_1} \leftrightarrow {\bf k -  k_1} $,
\begin{align}
    \frac{\partial P_{22}({\bf k} )}{\partial P(q)} \Delta q &= 2  \int_{{\bf k}_1} \left[ \frac{\partial P(k_1)}{\partial P(q)} P(|{\bf k} -{\bf k_1}|) + P(k_1) \frac{\partial P(|{\bf k} -{\bf k_1}|)}{\partial P(q)}  \right] Z_2^2({\bf k_1},{\bf k} -{\bf k_1})\Delta q \,,\nonumber\\
    &= 4  \int_{{\bf k}_1} \frac{\partial P(k_1)}{\partial P(q)} P(|{\bf k} -{\bf k_1}|) Z_2^2({\bf k_1},{\bf k} -{\bf k_1})\Delta q\,,\nonumber\\
    & = 4  \int_{{\bf k}_1}\delta_D(k_1-q) P(|{\bf k} -{\bf k_1}|) Z_2^2({\bf k_1},{\bf k} -{\bf k_1})\Delta q\nonumber\\
    & = 4 \;\frac{q^2 \Delta q}{(2\pi)^3}\int d\Omega_{\bf q} P(|{\bf k} -{\bf q}|) Z_2^2({\bf q},{\bf k} -{\bf q})\,,
\end{align}
where the last integral is performed over the solid angle of the ${\bf q}$ vector, $d\Omega_{\bf q}$.

On the other hand, for $P_{13}$ we get
\begin{align}
    \frac{\partial P_{13}({\bf k})}{\partial P(q)} & \!\Delta q = %\nonumber\\
    %&=
    6
    \Delta q\, Z_1({\bf k}) \!\!\int_{{\bf k}_1}\!\!\!\bigg[ \delta_D(k-q)  P(k_{1})Z_{3}(\mathbf{k},\mathbf{k}_{1},-\mathbf{k}_{1}) \!+\!
    P(k)   \delta_D(k_1-q) Z_{3}(\mathbf{k},\mathbf{k}_{1},-\mathbf{k}_{1})\bigg]\!,\nonumber\\
    & \!\!\!\! = 6
    \, Z_1({\bf k}) \bigg[ \Delta q \delta_D(k-q) \int_{{\bf k}_1} P(k_{1})Z_{3}(\mathbf{k},\mathbf{k}_{1},-\mathbf{k}_{1}) +    \frac{q^2 \Delta q}{(2\pi)^3} P(k)\int d\Omega_{\bf q} \,Z_{3}(\mathbf{k},\mathbf{q},-\mathbf{q})  \bigg]\,.
\end{align}

Finally, from the UV counterterms contribution we get,
\begin{align}
    \frac{\partial P^{\rm UV}({\bf k})}{\partial P(k)} = & -2\left(c_{0}k^{2}+c_{2} \beta \mu^{2}k^{2}+\tilde{c}b_1^6 \beta^{4}\mu^{4}k^{4}(1+\beta \mu^2)^2\right) \Delta q \delta_D(k-q)\,.\label{eq:UV-2-derivs}
\end{align}

\subsection{Derivative with respect to $\beta$}

In the case of scale-independent $\beta$, taking derivatives with respect to this parameter is straightforward. If we consider scale-dependent $\beta(k)$ it requires  more lengthy computation.
We start from the linear contribution,
\begin{equation}
\Delta q\frac{\partial P^{\rm lin}({\bf k})}{\partial \beta(q)}= 2 Z_1({\bf k}) P(k) b_1 \mu^2  \Delta q \,\delta_D(k-q) \,.
\end{equation}
For $P_{22}$ we have
\begin{align}
    \Delta q\left.\frac{\partial P_{22}({\bf k})}{\partial \beta(q)}\right|_{\beta(q)=\beta} = & \; 4\int_{{\bf k}_1}   Z_{2}(\mathbf{k}_1,\mathbf{k}-\mathbf{k}_1) \Delta q\left. \frac{\partial  Z_{2}(\mathbf{k}_1,\mathbf{k}-\mathbf{k}_1)}{\partial \beta(q)}\right|_{\beta(q)=\beta} \,P(q)P(|\mathbf{k}-\mathbf{k}_1|)\nonumber\\
    = & \; 4 \Delta q \delta(k-q) \int_{{\bf k}_1} Z_{2}(\mathbf{k}_1,\mathbf{k}-\mathbf{k}_1)Z_{2}^{\beta\beta}(\mathbf{k}_1,\mathbf{k}-\mathbf{k}_1)P(k_{1})P(|\mathbf{k}-\mathbf{k}_{1}|)\nonumber\\
    & +4 \frac{q^{2}\Delta q}{(2\pi)^3} P(q)\int d \Omega_{\bf q} P(|{\bf k} -{\bf q}|)Z_{2}(\mathbf{q},\mathbf{k}-\mathbf{q})Z_{2}^{\beta0}(\mathbf{q},\mathbf{k}-\mathbf{q})  \,,
\end{align}
where
\begin{align}
Z_{2}^{\beta\beta}(\mathbf{k}_1,\mathbf{k}_2) & =b_1 \mu^{2}G_{2}(\mathbf{k}_{1},\mathbf{k}_2)+b_1^{2}\frac{\mu k}{2}\left[\frac{\mu_{1}}{k_{1}}(1+\beta\mu_{2}^{2})+\frac{\mu_{2}}{k_2}(1+\beta\mu_{1}^{2})\right]\,,\\
Z_{2}^{\beta0}(\mathbf{k}_1,\mathbf{k}_2) & =b_1^{2} \beta \mu k\frac{\mu_{2}}{k_2}\mu_{1}^{2}\,.
\end{align}
For $P_{13}$, we get
\begin{align}
    \Delta q\left.\frac{\partial P_{13}({\bf k})}{\partial \beta(q)}\right|_{\beta(q)=\beta} = \; &  6 \Delta q\delta_D(k-q) \mu^2 b_1 P(k)
    \int_{\bf p}  \,\left.  Z_3(\mathbf{k},\mathbf{p},\mathbf{-p})\right|_{\beta(q)=\beta}\, P(p)\nonumber\\
    +&6 \,Z_1({\bf k})P(k)
    \int_{\bf p} \Delta q \,\left. \frac{\partial Z_3(\mathbf{k},\mathbf{p},\mathbf{-p})}{\partial \beta(q)}\right|_{\beta(q)=\beta}\, P(p)\,,
\end{align}
with
\begin{align}
    \Delta q & \, \left.
    \frac{\partial Z_3(\mathbf{k},\mathbf{p},\mathbf{-p})}{\partial \beta(q)}\right|_{\beta(q)=\beta}=\nonumber\\
    &\,\Delta q\,\delta_D(k-q) \bigg\{ b_1\mu^2G_3(\mathbf{k},\mathbf{p},\mathbf{-p})
    +\frac{1}{3}\mu k b_1 \bigg[\frac{\mu_{\bf p}}{p}\left(F_2(-\mathbf{p},\mathbf{k})-F_2(\mathbf{p},\mathbf{k})\right) \nonumber\\
    &+b_1\beta \left(\mu_{\mathbf{k-p}}^2 G_2(-\mathbf{p},\mathbf{k})- \mu_{\mathbf{k+p}}^2 G_2(\mathbf{p},\mathbf{k}) \right)\nonumber\\
    & + b_1\left(1+\beta \mu_{\mathbf p}^2\right)\left( \frac{\mu_{\mathbf{k-p}}}{|\mathbf{k-p}|} G_2(-\mathbf{p},\mathbf{k})+\frac{\mu_{\mathbf{k+p}}}{|\mathbf{k+p}|} G_2(\mathbf{p},\mathbf{k})\right)\bigg]\nonumber\\
    & - \frac{1}{3} b_1^3\mu^2 k^2 \beta \frac{\mu_{\mathbf{p}}^2}{p^2} (1+\beta \mu^2) -\frac{1}{6}b_1^3\mu^4 k^2\beta^2 \frac{\mu_{\mathbf{p}}^2}{p^2}\bigg\}\nonumber\\
    &+\frac{1}{3} \mu k b_1 \beta \Delta q
    \bigg\{ b_1 \frac{\mu_{\mathbf{p}}}{p} \bigg( \delta_D(|\mathbf{k-p}|-q) \mu_{\mathbf{k-p}}^2 G_2(-\mathbf{p},\mathbf{k})\nonumber\\
   & -\delta_D(|\mathbf{k+p}|-q)  \mu_{\mathbf{k+p}}^2 G_2(\mathbf{p},\mathbf{k} ) \bigg) \nonumber\\
    &+ \delta_D(p-q) b_1 \mu_{\mathbf{p}}^2\bigg(\frac{\mu_{\mathbf{k-p}}}{|\mathbf{k-p}|} G_2(-\mathbf{p},\mathbf{k})  + \frac{\mu_{\mathbf{k+p}}}{|\mathbf{k+p}|} G_2(\mathbf{p},\mathbf{k})  \bigg)\bigg\}\,,
\end{align}
where, compared to (\ref{z3}), $Z_3$ has been symmetrized and computed on the momenta configuration relevant for $P_{13}$.

\subsection{Derivatives with respect to bias parameters}

The bias parameters depend only on $z$, so the derivatives  do not offer any difficulty, being ordinary derivatives of the kernels $Z_2,Z_3$.
So for every $b_{i}$ we obtain
\begin{align}
    \frac{\partial}{\partial b_{i}}P_{gg}  = & \; 4\int_{\bf q} P(q)P(|\mathbf{k}-\mathbf{q}|)Z_{2}\left(\frac{\partial}{\partial b_{i}}Z_{2}\right) \nonumber \\
    & +6 Z_1({\bf k})P(k)\int_{\bf q} P(q)\left(\frac{\partial}{\partial b_{i}}Z_{3}\right) + \frac{\delta_{i1}}{b_1} P_{13} \,.
\end{align}

\subsection{Derivatives with respect to counterterms}

Also the derivatives with respect to the counterterm parameters $c_0,\, c_{2}$ and $\tilde{c}$ are trivial and there is no need to write them explicitly.

\begin{figure}
    \centering
    \includegraphics[width=.90\columnwidth]{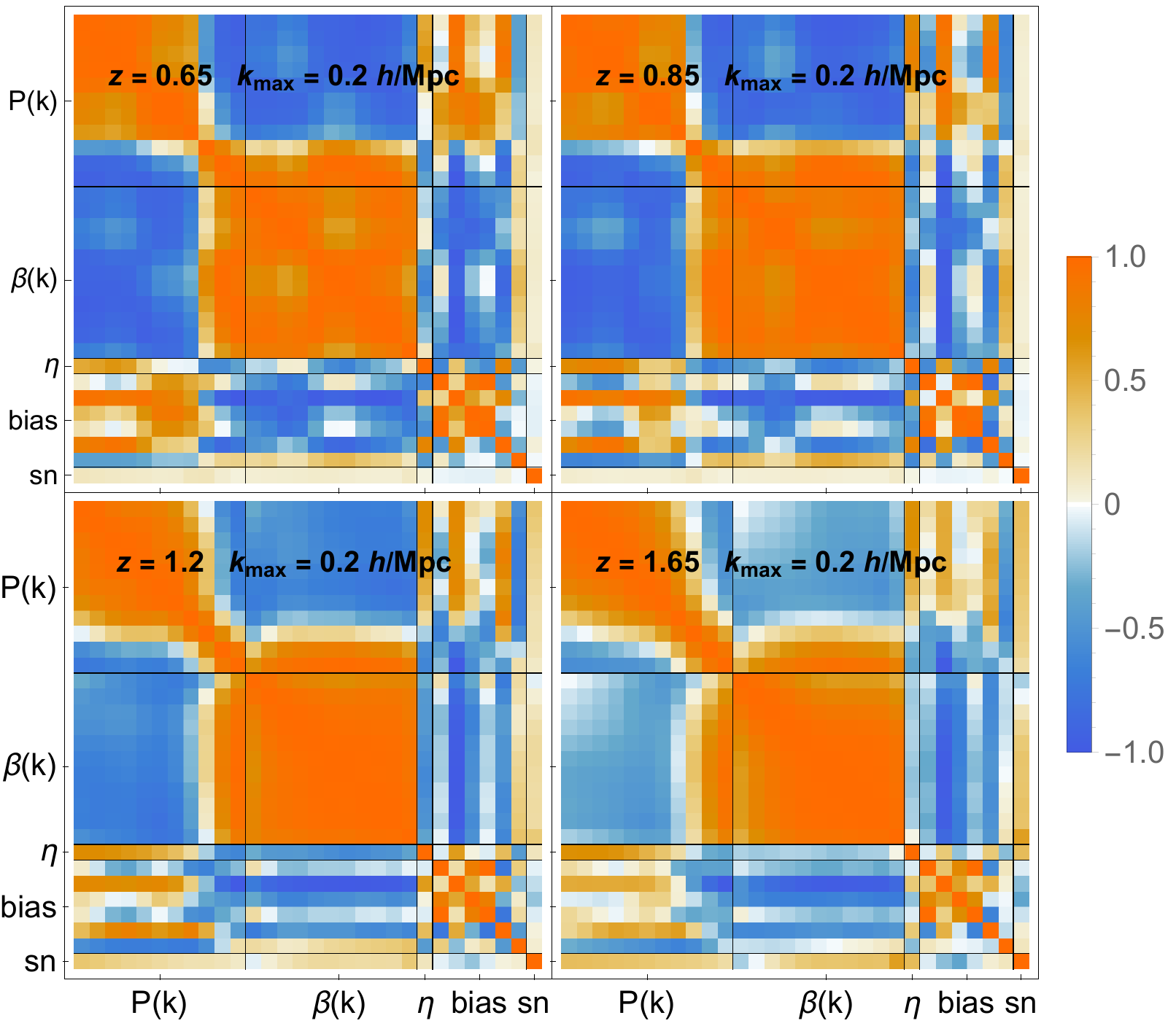}
    \caption{
    Correlation matrix for $k_{\rm max} = 0.2\,h$/Mpc for four redshift bins of DESI (upper panels) and Euclid (lower panels). The color scale depicts the correlation coefficient in each entry. The bias parameters are, in order: $\{c_0,\,\log b_1,\,b_2,\,b_{{\cal G}_2},\,b_{\Gamma_3},\, \log \sigma_f \}$. The shot-noise parameter $P_{\rm shot}$ is denoted by  ``sn''.
    \label{fig:corrmat}}
\end{figure}

\section{Correlations among parameters} \label{app:corr}

Figure~\ref{fig:corrmat} depicts the full correlation matrix between all parameters in our Fisher matrix for some redshift bins of  DESI (upper panels) and Euclid (lower panels) for $k_{\rm max} = 0.2 \, h$/Mpc. For better clarity in the plots, in this appendix we adopt only 11 $k$ bins instead of 20, which are ordered from higher $k$ to lower $k$. As can be seen the $\beta(k)$ parameters are highly correlated among themselves in all bins. The $P(k)$ bins are instead correlated with the neighbouring bins but anti-correlated with faraway bins. These high correlations are also related to the large marginalized errors in the individual $P(k)$ and $\beta(k)$, as illustrated by Figure~\ref{fig:delta-logP-logB}.

\begin{figure}
    \centering
    \includegraphics[width=.57\columnwidth]{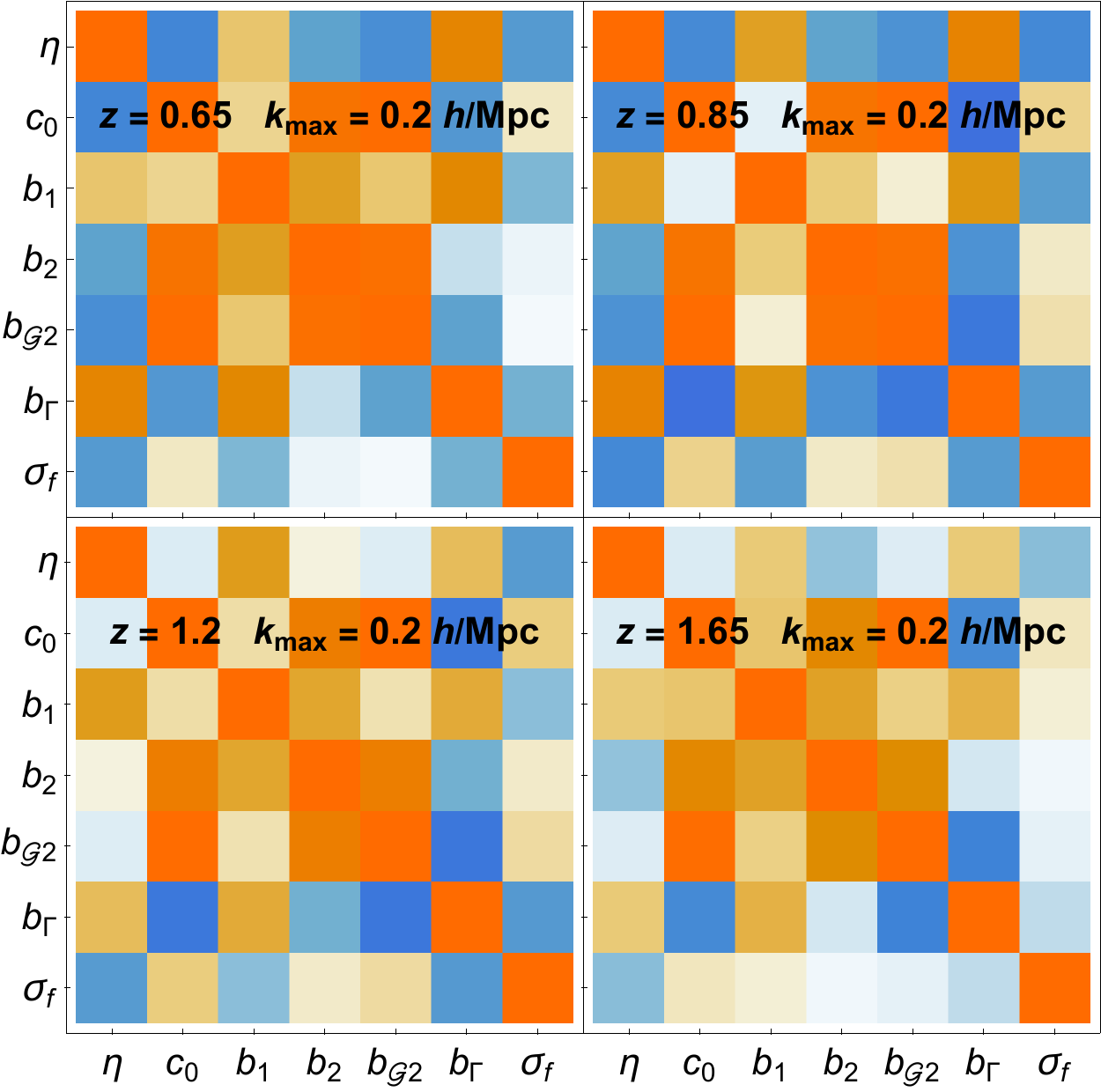}\quad\;
    \raisebox{.0018\height}{\includegraphics[width=.38\columnwidth]{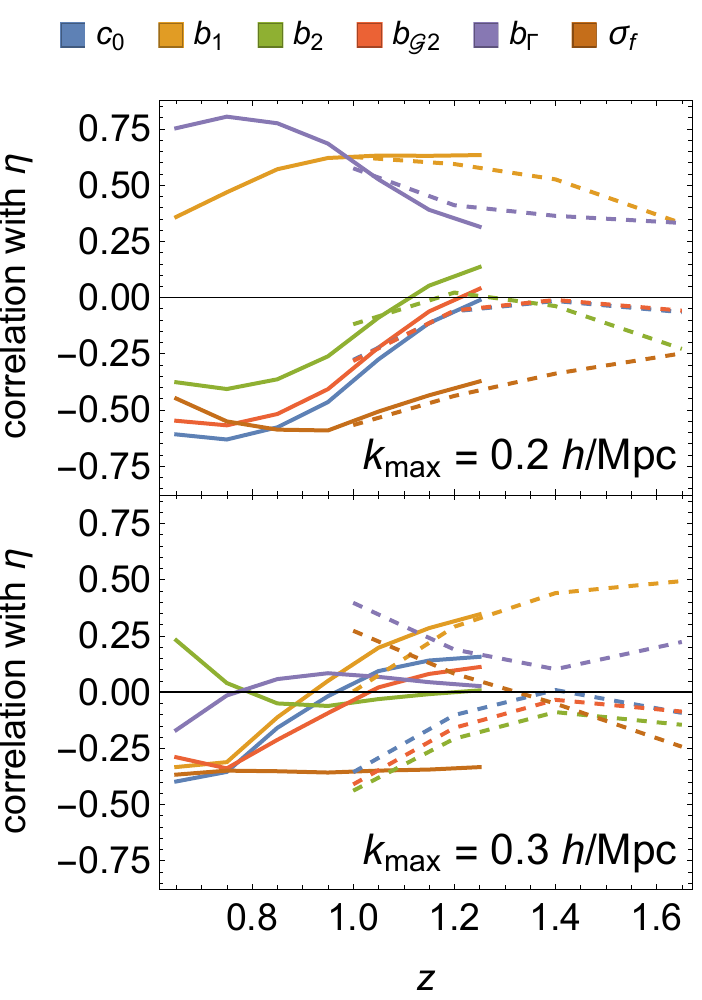}}
    \caption{
    [Left] Same as Figure~\ref{fig:corrmat}, but zooming in $\eta$, bias, counterterm and FoG parameters. [Right] Correlation between $\eta$ and the other parameters as a function of redshift for the same $k_{\rm max} = 0.2\,h/$Mpc (upper panel) and also for $k_{\rm max} = 0.3\,h/$Mpc (bottom panel). Solid (dashed) lines represent the DESI (Euclid) forecasts.
    \label{fig:corrmatzoom}}
\end{figure}

Figure~\ref{fig:corrmatzoom} shows a close-up of the sub-matrix in $\eta$ and the different non-linear bias parameters. The right panel also depicts how the correlations of $\eta$ and the other five parameters evolve with redshift. For $k_{\rm max} = 0.2 \, h$/Mpc there are strong correlations between $\eta$ and $b_{\Gamma}$ (positive) and both $b_{{\cal G}_2}$ and $c_0$ (negative). These correlations diminish at higher $z$, although the one with $b_{\Gamma}$ remains significant. For $k_{\rm max} = 0.3 \, h$/Mpc, $\eta$ is much less correlated with these nuisance parameters, and therefore less sensitive to systematics in them.

\section{Scaling with $k_{\rm max}$ of different parameters}\label{app:scaling}

\begin{figure}
    \centering
    \includegraphics[width=.47\columnwidth]{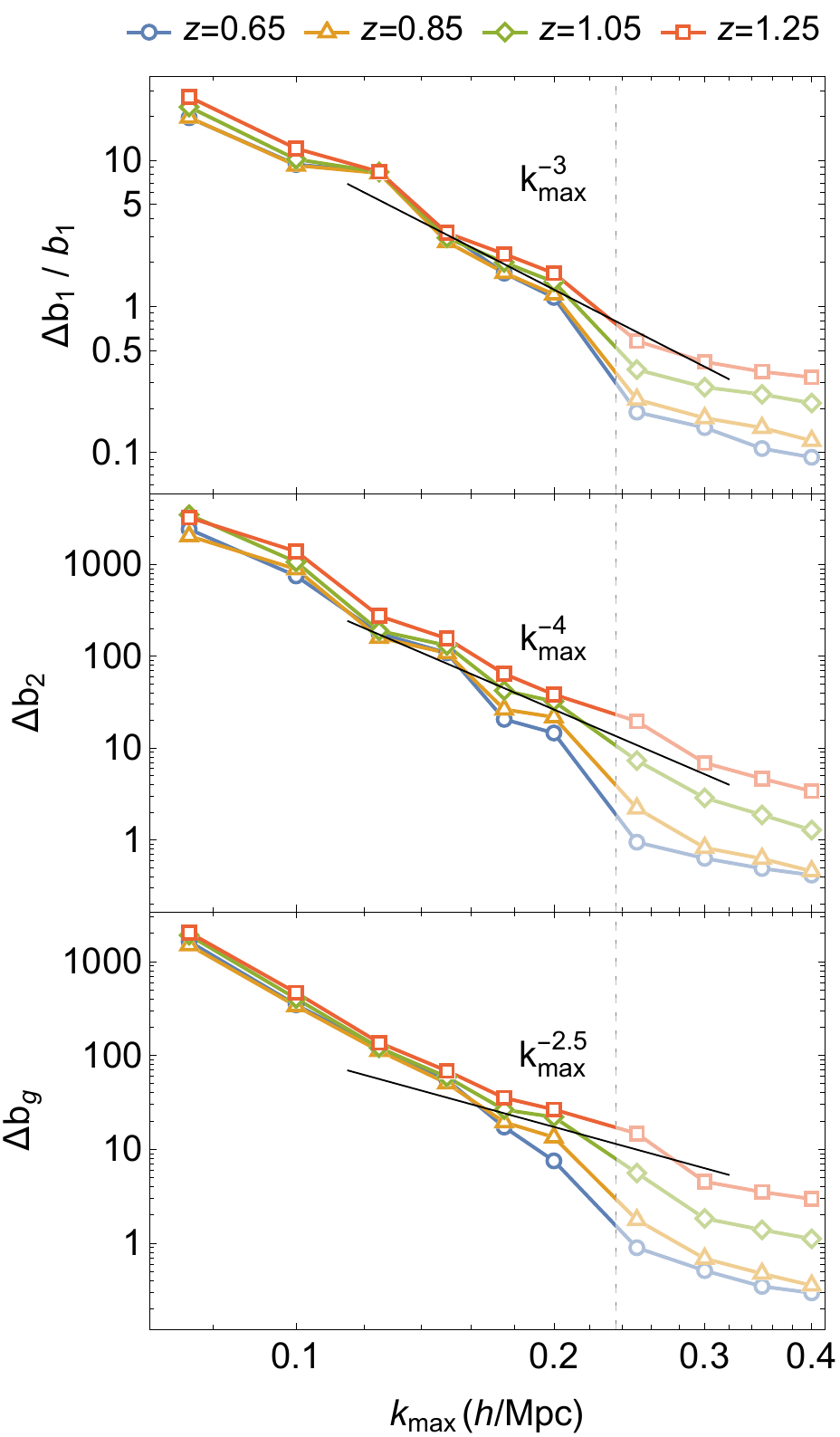}\;\;\;
    \includegraphics[width=.47\columnwidth]{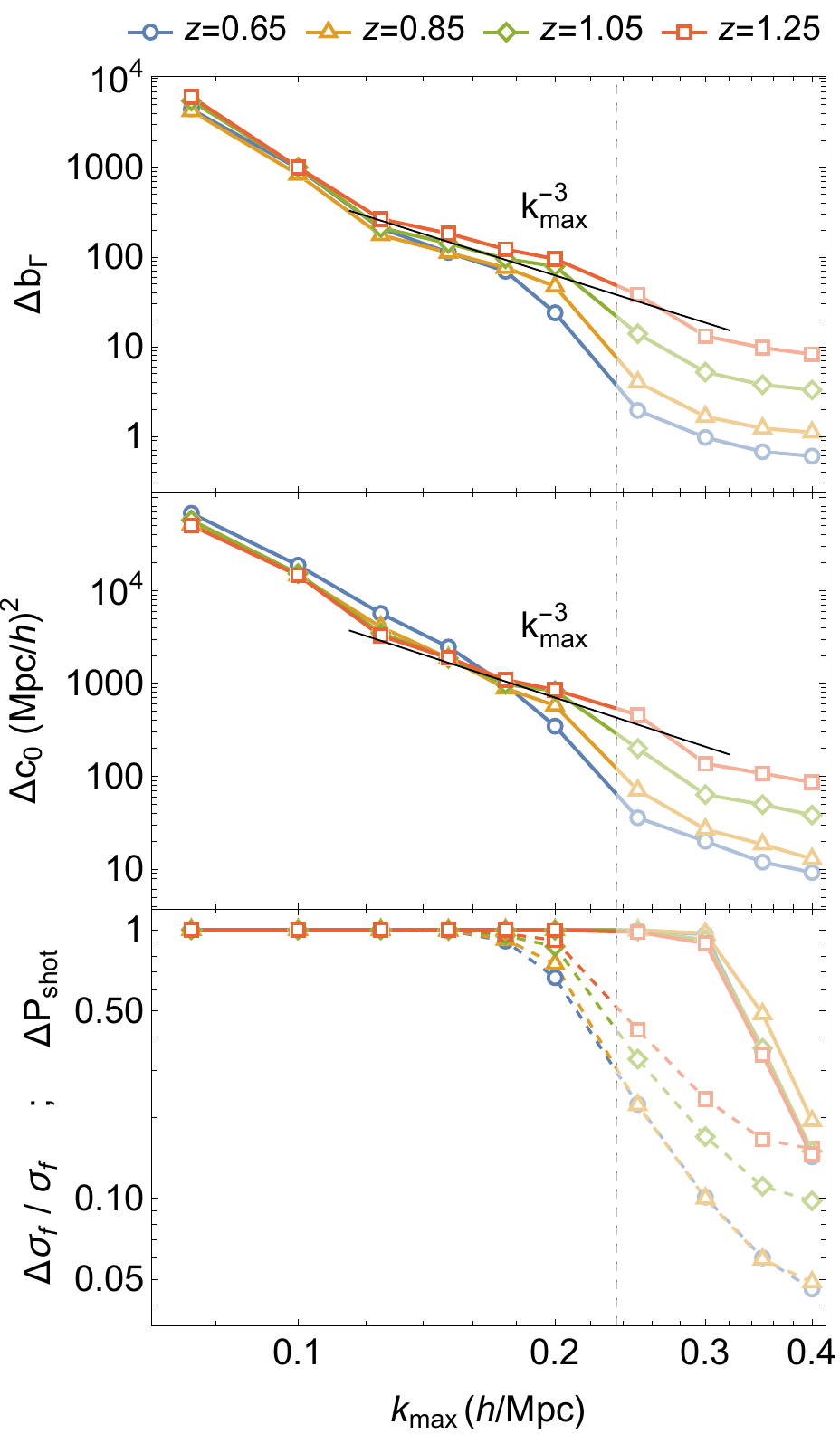}
    \caption{
    Similar to Figure~\ref{fig:delta-eta-scaling}, but for all of the nuisance parameters. In the bottom right panel, we show both FoG parameters, which are the only ones with informative priors; $\log \sigma_f$ ($P_{\rm shot}$) in dashed (full) lines.
    \label{fig:scaling-eta-others}}
\end{figure}

\begin{figure}
    \centering
    \includegraphics[width=.47\columnwidth]{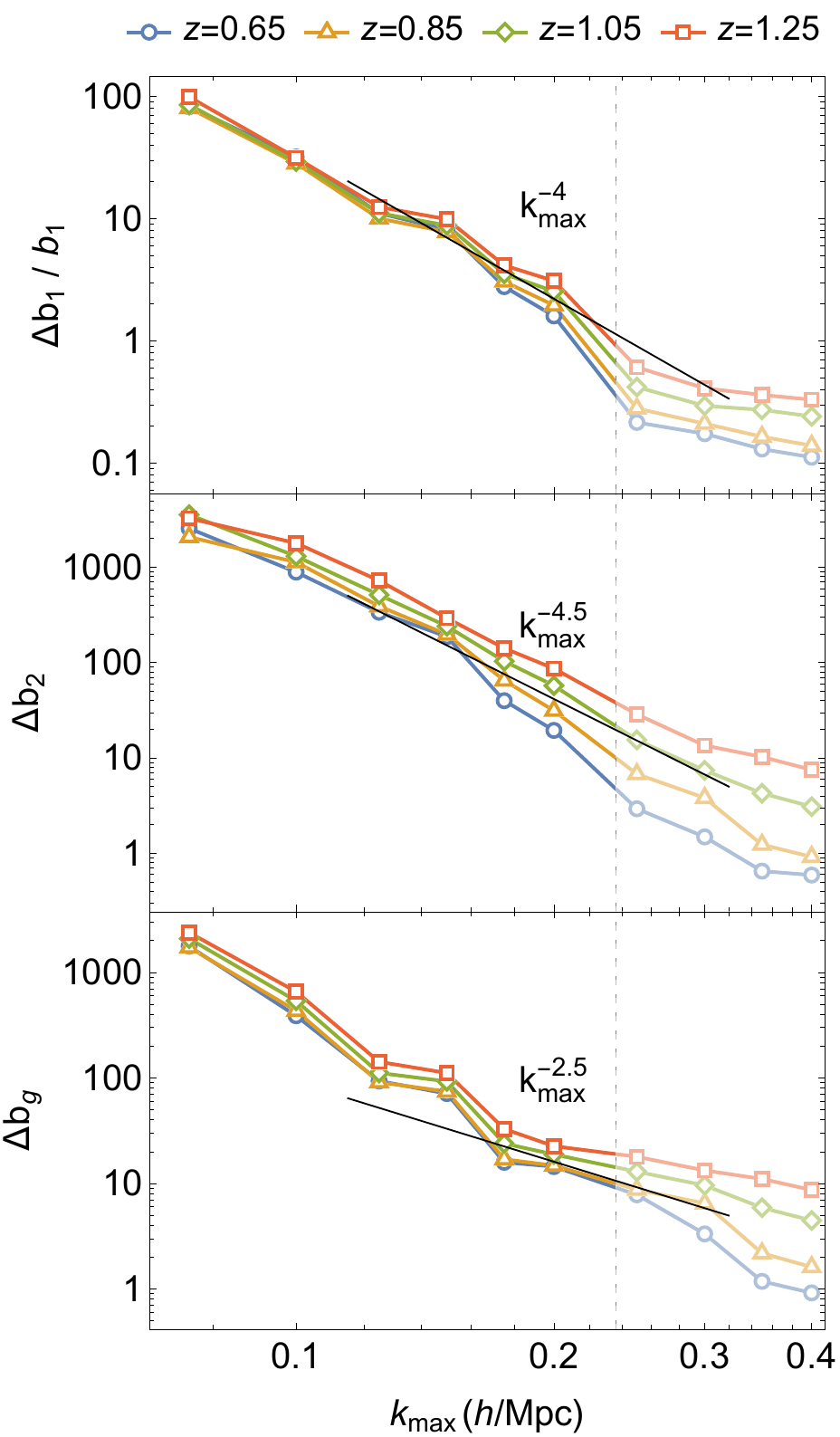}\;\;\;
    \includegraphics[width=.47\columnwidth]{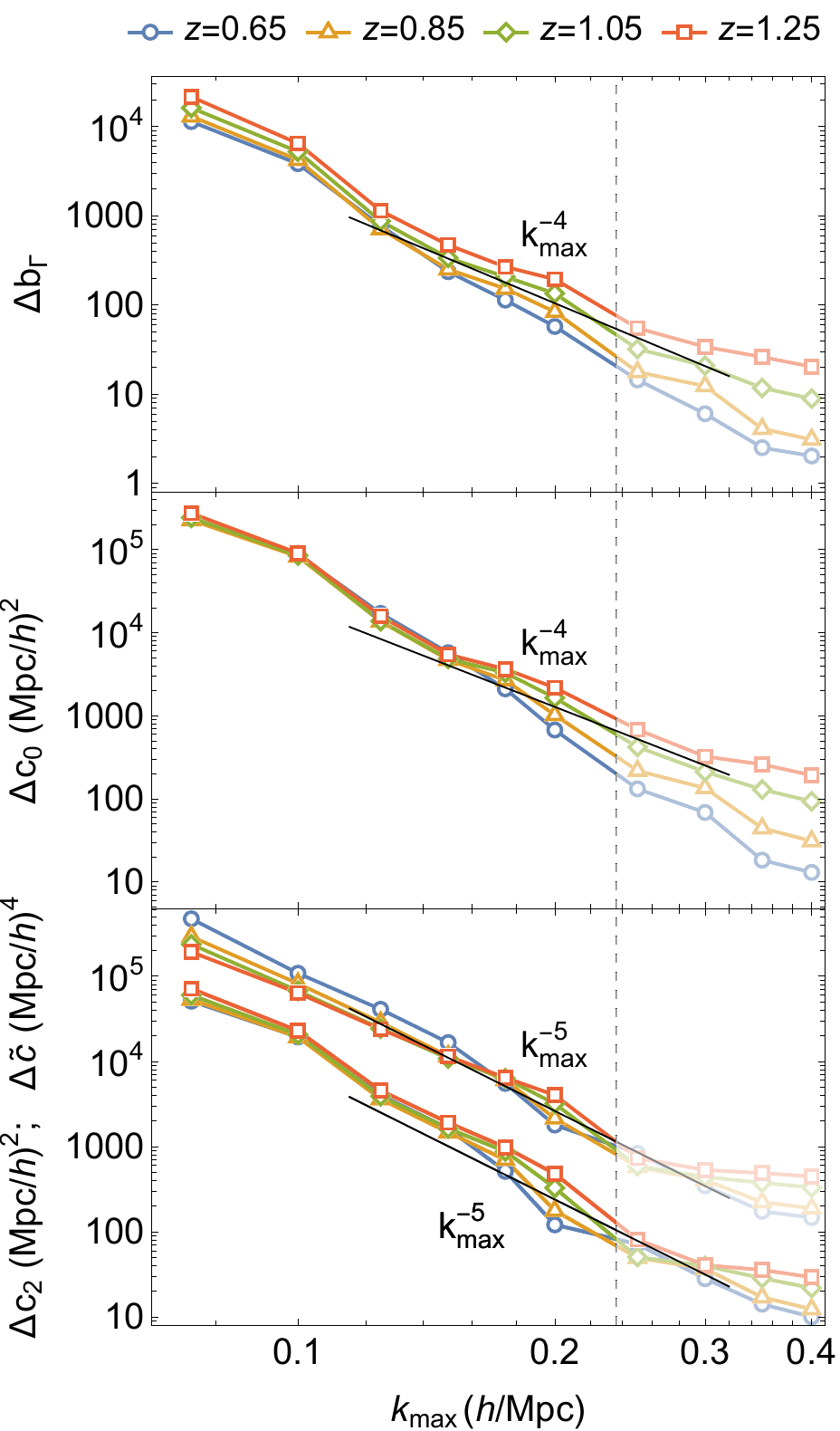}
    \caption{
    Same as Figure~\ref{fig:scaling-eta-others}, but for the case in which the counterterms $c_2$ and $\tilde{c}$ are added with uninformative priors and the FoG parameter $\sigma_f$ kept fixed.
    \label{fig:scaling-eta-others-c2ctilde}}
\end{figure}

In this Appendix we extend to the different nuisance parameters the analysis of how the precision scales as a function of the cut-off scale $k_{\rm max}$, depicted for $\eta$ in Figure~\ref{fig:delta-eta-scaling}. Figures~\ref{fig:scaling-eta-others} and~\ref{fig:scaling-eta-others-c2ctilde} summarize the results for DESI in the free $\beta(k)$ case with uninformative priors: the former for the exponential FoG modelling, the latter for the FoG modelling using instead the parameters $c_2$ and $\tilde{c}$.  As can be seen, in both cases the results are comparable, and obey similar power laws in the range $k_{\rm max}\le 0.3 h/$Mpc. The largest difference being a somewhat larger redshift dependence for the exponential FoG case in the range $k_{\rm max} > 0.2 h/$Mpc.

We also remark that while $\eta$ scales with an exponent close to $-1.5$, the different bias and counterterm parameters exhibit a steeper dependence on $k_{\rm max}$, typically with exponents around $-3$, except for $b_2$ which scales with exponent close to $-1.5$, the different bias and counterterm parameters exhibit a steeper dependence on $k_{\rm max}^{-4}$. $P_{\rm shot}$ instead is prior-dominated in all cases, even though the prior used is conservative from the physical point of view.

\section{AP effect and the band-power marginalization }\label{app:band-power}

In our approach, both $P(k)$ and $\beta(k)$ are marginalized over in band-powers. This might lead one to
wonder whether the $\eta$ dependence of the AP effect is erased. Here we show by a simple analytical argument why this is not the case. For simplicity, we consider just a single $k$-bin, centered on $\hat k$, and $n_b$ $\mu$-bins. The binned power spectrum is then given by
\begin{align}
    \bar P_i \equiv \frac{1}{\Delta \mu} \int_{\mu_i-\Delta \mu}^{\mu_i+\Delta \mu}\dd \mu \,P(\hat k,\mu)\,,
\end{align}
where $\Delta \mu =1/(2 n_b)$, $\mu_i=(2 i+1)\Delta \mu $ and $i=0,\cdots, n_b-1$, and we consider only bins with $0\le \mu_i\le 1$, exploiting the parity property $P(k,-\mu)=P(k,\mu)$.

The Fisher matrix element corresponding to a set of parameters $\theta$ is given by
\begin{equation}
    F_{\alpha \beta}=  V V_{\hat k}\, \Delta \mu  \sum_{i=0}^{n_b-1}F^i_{\alpha \beta}\,,
    \label{fullF}
\end{equation}
where
\begin{equation}
    F^i_{\alpha\beta}= \frac{\partial\log \bar P_i}{\partial\theta_\alpha}\frac{\partial\log \bar P_i}{\partial\theta_\beta}\,.
\end{equation}
We assume a linear power spectrum without shot noise,
\begin{equation}
    P(\hat k,\mu) = \left(1+\beta(\hat k) \mu^2\right)^2 P_0(\hat k)\,,
\end{equation}
and the set of parameters $\theta=\{\log P_0(\hat k),\log \beta(\hat k),\log\eta\}$. We then compute the derivatives
\begin{align}
    \frac{\partial\log \bar P_i }{\partial \log P_0(\hat k)}& =1\,, \nonumber\\
    \frac{\partial\log \bar P_i }{\partial \log \beta(\hat k)}&=
    \frac{\partial \log F[\beta(\hat k),\mu_i,\Delta \mu]}{\partial \log \beta(\hat k)}\,,\nonumber\\
    \frac{\partial\log \bar P_i }{\partial \log \eta}&=\frac{1}{\bar P_i} \frac{1}{\Delta \mu} \int_{\mu_i-\Delta \mu}^{\mu_i+\Delta \mu}\dd \mu \,\bigg[ \frac{\partial P(\hat k,\mu) }{\partial \log \hat k}\mu^2+  \frac{\partial P(\hat k,\mu) }{\partial \log \hat \mu}(1-\mu^2)\bigg]\,,\nonumber\\
    &= n_k \,G[\beta(\hat k),\mu_i,\Delta \mu] + h_k\, I[\beta(\hat k),\mu_i,\Delta \mu]+H[\beta(\hat k),\mu_i,\Delta \mu]\,,
\end{align}
where
\begin{align}
    F[\beta(\hat k),\mu_i,\Delta \mu]&\equiv  \frac{1}{\Delta \mu} \int_{\mu_i-\Delta \mu}^{\mu_i+\Delta \mu} \dd \mu \, (1+\beta(\hat k) \mu^2)^2\,,\nonumber\\
    G[\beta(\hat k),\mu_i,\Delta \mu]&\equiv \frac{1}{F[\beta(\hat k),\mu_i,\Delta \mu]} \frac{1}{\Delta \mu} \int_{\mu_i-\Delta \mu}^{\mu_i+\Delta \mu}\dd \mu \, (1+\beta(\hat k) \mu^2)^2\, \mu^2\,,\nonumber\\
    I[\beta(\hat k),\mu_i,\Delta \mu]&\equiv \frac{2}{F[\beta(\hat k),\mu_i,\Delta \mu]} \frac{1}{\Delta \mu} \int_{\mu_i-\Delta \mu}^{\mu_i+\Delta \mu}\dd \mu \, (1+\beta(\hat k) \mu^2)\, \mu^4\,,\nonumber\\
    H[\beta(\hat k),\mu_i,\Delta \mu]&\equiv \frac{4 \beta(\hat k)}{F[\beta(\hat k),\mu_i,\Delta \mu]} \frac{1}{\Delta \mu} \int_{\mu_i-\Delta \mu}^{\mu_i+\Delta \mu}\dd \mu \, (1+\beta(\hat k) \mu^2)\, \mu^2(1-\mu^2)\,,
\end{align}
and
\begin{equation}
    n_k\equiv\frac{\partial \log P_0(\hat k)}{\partial \log \hat k}\,,\qquad
    h_k\equiv\frac{\partial \log \beta(\hat k)}{\partial \log \hat k}\,.
\end{equation}
Then we have the matrix
\begin{equation}
    F^i_{\alpha \beta}=
    \left(\begin{array}{ccc}
    1 & \frac{\partial \log F_i}{\partial \log \beta} & n_k G_i+h_k I_i+ H_i\\
    \frac{\partial \log F_i}{\partial \log \beta} & \left(\frac{\partial \log F_i}{\partial \log \beta}\right)^2 & \left(\frac{\partial \log F_i}{\partial \log \beta}\right)\left( n_k G_i+h_k I_i+ H_i\right)\\
    n_k G_i+h_k I_i+ H_i & \left(\frac{\partial \log F_i}{\partial \log \beta}\right)\left( n_k G_i+h_k I_i+ H_i\right) & \left( n_k G_i+h_k I_i+ H_i\right)^2
    \end{array}\right)\,.
\end{equation}
This matrix is singular, because any AP change in $k$, evaluated in a single $\mu$ bin, can be compensated by a change in $P_0(k)$. Also the sum of two matrices evaluated at two different $\mu_i$'s is singular, due to the extra freedom in $\beta(\hat k)$. Indeed, when computing the full matrix (\ref{fullF}), one finds that it is singular for $n_b=1$ and $n_b=2$, but becomes non singular for $n_b\ge 3$. This can be understood as follows. The AP effect amounts to a quadrupolar distortion, which affects all the power spectrum multipoles. In our approach, the effects on the power spectrum monopole and quadrupole can be reabsorbed by changing the power spectrum amplitude and $\beta$, and therefore we need at least three multipoles to detect it. If instead of multipoles we compute $\mu$-bins,  we need at least three independent angular information (three bins) to detect the AP distortion.

Of course, the larger the number of bins the better the AP effect is recovered. Evaluating the determinant of the matrix  (\ref{fullF}), we see that the asymptotic limit (corresponding to infinite number of bins) is in practice  approached for $n_b$ about 10. This is the assumption  in the computations in this paper, where, by integrating in $\mu$ as in Eq.~(\ref{FMi}), we implicitly take the $n_b\to \infty$ limit.

Notice also that the effect depends on the local slopes $n_k$ and $h_k$  but not to the exact location of the wiggles. Some location-dependent sensitivity remains, though, because of the different weight introduced by $V_{\hat k}$ when one considers more $k$-bins. Inclusion of shot noise, several $k$-bands, and non-linear corrections will not change qualitatively this conclusion.

One can also wonder what happens to the information on $\eta$ with increasing the number of $k$ bins. Fixing for simplicity all parameters, including $\beta(k)$, but not $P(k)$ or $\eta$, one finds, in the same simplified setting above and integrating over $\mu$, that for a generic number of bins  the relative error on $\eta$ becomes
\begin{equation}
    \sigma_\eta ^2=\frac{45}{8V(\sum_i n_{k_i}^2 V_i)}\,,
\end{equation}
where $V_i=(2\pi)^{-2}k_i^2\Delta k_i$. Approximating
\begin{equation}
    \sum_i n_{k_i}^2 V_i \,=\, \frac{1}{(2\pi)^{2}} \int n(k)^2 k^2 \dd k \,,
\end{equation}
one sees that the constraints on $\eta$ are asymptotically independent of the $k$-binning, as we confirmed numerically in the main text.

These arguments  show that our model-independent approach relies essentially on the assumption of statistical isotropy.

\bibliographystyle{JHEP2015}

\bibliography{references,scaling_bib}
\label{lastpage}
\end{document}